\documentclass[12pt]{article}

\usepackage{jheppub}

\usepackage{tensor}

\newcommand{\op}{\mathcal{O}}
\newcommand{\vep}{\varepsilon}
\newcommand{\tr}{\operatorname{Tr}}
\newcommand{\dt}{\frac{d}{2}}
\newcommand{\D}{\Delta}
\newcommand{\pd}{\mathcal{D}}

\newcommand{\eul}{\gamma_E}

\newcommand{\vev}[1]{\langle #1 \rangle}
\newcommand{\fvev}[1]{\left\langle #1 \right \rangle}
\newcommand{\mvev}[2]{#1\langle #2 #1\rangle}
\newcommand{\logp}[1]{\log\left( #1 \right)}

\newcommand{\beq}{\begin{equation}}
\newcommand{\eeq}{\end{equation}}

\begin{document}
\title{Entanglement entropy of excited states  in conformal perturbation
theory and the Einstein equation} 

\author{Antony J. Speranza}
\affiliation{Maryland Center for Fundamental Physics, University of Maryland, College Park, 
MD 20742, USA. }
\emailAdd{asperanz@umd.edu}

\date{Jan 11, 2016}
\abstract{
For a conformal field theory (CFT) deformed by a relevant operator, the entanglement entropy
of a ball-shaped region may be computed as a perturbative expansion in the  
coupling.  A similar perturbative expansion exists for excited states near the vacuum.  Using
these expansions, this work investigates the behavior of excited state entanglement entropies
of small, ball-shaped regions.   
The motivation for these calculations is Jacobson's recent 
work on the equivalence of the Einstein equation and the hypothesis of 
maximal vacuum entropy [arXiv:1505.04753], 
which relies on a conjecture stating that the behavior of these 
entropies is sufficiently similar to a CFT.   
In addition to the expected type of terms which scale with the ball radius as $R^d$, 
the entanglement entropy calculation gives
rise to terms scaling as $R^{2\D}$, where $\D$ is the dimension of the deforming operator.
When $\D\leq\dt$, the latter terms dominate the former, and suggest that a modification to 
the conjecture is needed.
}

\maketitle
\flushbottom

\section{Introduction}

Entanglement entropy is a quantity with many profound and surprising
connections to spacetime geometry, and is suspected to play an important role in
a complete description of quantum gravity.  It has featured prominently  explanations
of the origin of black hole entropy \cite{Bekenstein1973, Hawking1975,
Sorkin:2014kta,Bombelli1986,Frolov1993,Srednicki1993, Solodukhin2011}, stemming
from the similarity between the area law for the Bekenstein-Hawking entropy and the 
area law for entanglement entropy.  
In holographic theories, the entanglement
entropy of the CFT is intimately related to the  bulk geometry by virtue of the
Ryu-Takayanagi (RT) formula \cite{Ryu:2006bv, Ryu:2006ef} and its covariant generalization
\cite{Hubeny2007}, which state that the entropy 
is dual to the area of an extremal surface in the bulk.
These connections motivate the compelling idea that spacetime geometry and its
dynamics may emerge from the entanglement structure of quantum fields.   
This ``geometry from entanglement'' program has recently found a concrete realization
in holography, where the bulk linearized Einstein equations were shown to
follow from
the RT formula \cite{Lashkari2014,Faulkner:2013ica, Swingle2014}.

Another recent development is a proposal by Jacobson \cite{Jacobson2015}, 
which builds upon his original derivation of the Einstein equation as a thermodynamic 
equation
of state \cite{Jacobson1995}.  In this new work, he postulates that the local quantum gravity
vacuum is an equilibrium state, in the sense that it is a state of maximal entanglement 
entropy.  It is then demonstrated that this hypothesis is equivalent to the Einstein equation.
Entanglement entropy is the key object relating
the geometrical quantities
on the one hand to the stress-energy of matter fields on the other.  In this case, the connection
between entanglement entropy and geometry stems from the area law;
the entropy is dominated by modes near the entangling surface, and hence scales as 
the area  \cite{Srednicki1993}.
On the other hand, it relates to matter stress-energy through the modular Hamiltonian, which, 
for a ball-shaped region in a CFT vacuum, 
is constructed from the stress-energy
tensor.  

The ability to express the modular Hamiltonian of a ball in terms of a simple integral of 
the stress tensor is special
to a CFT.  Extending the argument for the equivalence between Einstein's equations and 
maximal vacuum entanglement  to  non-conformal  fields
requires taking the  ball to be much smaller than any length scale appearing in the 
field theory.  Since the theory will flow to an ultraviolet (UV) fixed point at 
short length scales, one expects to recover CFT behavior in this limit.
Jacobson made a conjecture about the form of the entanglement entropy for excited states
in small spherical regions that allowed the argument to go through.  
The purpose of the present paper is to check this conjecture 
using conformal perturbation theory (see also \cite{Carroll2016} for 
alternative ideas for checking the conjecture).  

In this work, we will consider a CFT deformed by a relevant operator $\op$ of dimension $\D$, 
and examine the
entanglement entropy for a class of excited states formed by a path integral over
Euclidean space.  The entanglement entropy in this case may be evaluated using recently
developed perturbative techniques \cite{Rosenhaus2014, Rosenhaus2014a, 
Rosenhaus:2014ula,Rosenhaus:2014zza,Faulkner2015,Faulkner2015a} which express
the entropy in terms of correlation functions, and notably do not rely on the replica 
trick \cite{Susskind:1993ws, Callan1994}.  In particular, one knows from the expansion
in \cite{Rosenhaus2014,Rosenhaus:2014ula} that the first correction to the 
CFT entanglement entropy comes from the $\op\op$ two-point function and the
$K\op\op$ three point function, where $K$ is the CFT vacuum modular Hamiltonian.  
However, those works did not account for the noncommutativity of the density matrix
perturbation $\delta\rho$ with the original density matrix $\rho_0$, so the results
cannot be directly applied to find the finite change in entanglement entropy between the perturbed
theory excited state and the CFT ground state.\footnote{However, references 
\cite{Rosenhaus:2014ula,
Rosenhaus:2014zza} are able to reproduce universal logarithmic divergences when they
are present.  }  Instead, we will apply the technique developed by Faulkner \cite{Faulkner2015} 
to compute
these finite changes to the entanglement entropy, which we review in section
\ref{sec:EEballs}.  The result for the change in entanglement entropy between the excited
state and vacuum is
\beq  \label{eqn:Dneqdt}
\delta S =  \frac{2\pi\Omega_{d-2} }{d^2-1} \left[R^d\left(\delta\vev{T^g_{00}}  
-\frac{1}{2\D-d} \delta
\vev{T^g}  \right) - R^{2\D} \vev{\op}_g\delta\vev{\op} \frac{\D\Gamma(\dt+\frac32) 
\Gamma(\D-\dt+1)}{(2\D-d)^2 \Gamma(\D+\frac32)}  \right],
\eeq
which holds to first order in the variation of the state and for $\D\neq\dt$.  Here, $\Omega_{d-2}
= \frac{2\pi^{\dt-\frac12}}{\Gamma(\dt-\frac12)}$ is the volume of the unit $(d-2)$-sphere, $R$
is the radius of the ball,
$T^g_{\mu\nu}$ is the stress tensor of the deformed theory with trace $T^g$, $\vev{\op}_g$
stands for the vacuum expectation value of $\op$, and the $\delta$ refers to the change in 
each quantity relative to the vacuum value.  

The case $\D=\dt$ requires special attention, since the 
above expression degenerates at that value of $\D$.  The result for $\D=\dt$ is 
\beq\label{eqn:Deqdt} 
\delta S = 2\pi \frac{\Omega_{d-2}}{d^2-1} R^d \left[ \delta\fvev{T^g_{00}} + \delta \vev{T^g} 
\left(\frac2d-\frac12H_{\frac{d+1}{2}} + \log\frac{\mu R}{2} \right)-\dt \vev{\op}_g\delta\vev{\op}
\right],
\eeq
where $H_{\frac{d+1}{2}}$ is a harmonic number, defined for the integers by $H_n = 
\sum_{k=1}^n \frac1k$ and for arbitrary values of $n$ by $H_n = \eul+\psi_0(n+1)$ with
$\eul$ the Euler-Mascheroni constant, and $\psi_0(x) = \frac{d}{dx}\log\Gamma(x)$ the
digamma function.  
This result depends on a renormalization scale $\mu$ which arises due to an ambiguity 
in defining a renormalized value for the vev $\vev{\op}_g$.  The above result only superficially
depends on $\mu$, but this dependence cancels between the $\log\frac{\mu R}{2}$  and 
$\vev{\op}_g$ terms.
These results agree with recent holographic calculations \cite{Casini2016a}, and this 
work therefore establishes that those results extend beyond holography.

In both equations (\ref{eqn:Dneqdt}) and (\ref{eqn:Deqdt}), the first terms scaling as $R^d$ 
take the  form required for Jacobson's argument.  However, when $\D\leq\dt$, 
the terms scaling as $R^{2\D}$ or $R^d \log R$ dominate over this term in the small $R$ limit.  
This leads to some tension with the argument for the equivalence of the Einstein equation and 
the hypothesis of maximal vacuum entanglement.  We revisit this point in section
\ref{sec:EEimps} and suggest some possible resolutions to this issue.

Before presenting the calculations leading to equations (\ref{eqn:Dneqdt}) and 
(\ref{eqn:Deqdt}), we briefly review Jacobson's argument in section \ref{sec:EE=EE}, where
we describe in more detail the form of the variation of the entanglement entropy that would 
be needed for the derivation of the Einstein equation to go through.
We also provide a review of Faulkner's method for calculating
entanglement entropy in section \ref{sec:EEballs}, since it will be used heavily in the sequel.  
Section \ref{sec:excited} describes the type of excited states considered in this paper, including
an important discussion of the issue of UV divergences in operator expectation values.  
Following this, we present the derivation of the above result to first order in
$\delta\vev{\op}$ in section \ref{sec:harder}.  Finally, we discuss
the implications of these results for the Einstein equation derivation and avenues for 
further research in section \ref{sec:discussion}.

\section{Background}

\subsection{Einstein equation from entanglement equilibrium} \label{sec:EE=EE}
This section provides a brief overview of Jacobson's argument for
the equivalence of the Einstein equation and 
the maximal vacuum entanglement hypothesis
\cite{Jacobson2015}.  The hypothesis states that the entropy of a small  geodesic
ball is maximal in 
a vacuum configuration
of quantum fields coupled to gravity, i.e.\ the vacuum is an equilibrium state.  
This implies that as the state is varied at fixed volume away from 
vacuum, the change in the entropy must be zero at first order in the variation.  
In order for this to be possible, the entropy increase of the matter fields must be compensated
by an entropy decrease due to the variation of the geometry.  Demanding that these
two contributions to the entanglement entropy cancel leads directly to the Einstein equation. 

Consider the simultaneous variations of the metric and the state of the quantum fields, $(\delta
g_{ab}, \delta \rho)$.  
The metric variation induces a change $\delta A$ in the surface area of the geodesic ball,
relative to the surface area of a ball with the same volume in the unperturbed metric.
Due to the area law, this leads to a proportional change $\delta S_\text{UV}$
in the entanglement
entropy
\beq \label{eqn:dSUV}
\delta S_\text{UV} = \eta \delta A.
\eeq
Normally, the constant $\eta$ is divergent and regularization dependent; however, one further
assumes that quantum gravitational effects render it finite and universal.   For small enough
balls, the area variation is expressible in terms of the $00$-component of the Einstein tensor
at the center of the ball.  
Allowing for the background geometry from which the variation is taken to 
be any maximally symmetric space, with Einstein tensor $G_{ab}^\text{MSS} = -\Lambda
g_{ab}$,  (\ref{eqn:dSUV}) becomes \cite{Jacobson2015}
\beq \label{eqn:dSUVG}
\delta S_\text{UV}=  -\eta\frac{\Omega_{d-2} R^d}{d^2-1} (G_{00} + \Lambda g_{00}) .
\eeq

The variation of the quantum state produces the compensating contribution to the entropy. At
first order in $\delta \rho$, this is given by the change in the modular Hamiltonian $K$,
\beq \label{eqn:dSIR}
\delta S_\text{IR} = 2\pi \delta\vev{K},
\eeq
where $K$ is related to $\rho_0$, the reduced density matrix of the vacuum restricted to the ball, 
via
\beq \label{eqn:rho0}
\rho_0 = e^{-2\pi K}/Z, 
\eeq
with the partition function $Z$ providing the normalization.  Generically, $K$ is a complicated,
nonlocal operator; however, in the case of a ball-shaped region of a CFT, it is given by 
a simple integral of the energy density over the ball \cite{Hislop1982, Casini:2011kv},
\beq \label{eqn:K}
K = \int_\Sigma d\Sigma^a \zeta^b T_{ab} = \int_\Sigma d\Omega_{d-2} dr\, r^{d-2} \left(
\frac{R^2-r^2}{2R}\right) T_{00}.
\eeq
In this equation, $\zeta^a$ is the conformal Killing vector in Minkowski space\footnote{
The conformal Killing vector is different 
for a general maximally symmetric space \cite{Casini2016a}.  
However, the Minkowski space vector is sufficient as long as $R^2\ll\Lambda^{-1}$.} that fixes the 
boundary $\partial \Sigma$ of the ball.  With the standard Minkowski time $t=x^0$ and spatial
radial coordinate $r$, it is given by
\beq \label{eqn:CKV}
\zeta = \left(\frac{R^2-r^2-t^2}{2R}\right) \partial_t - \frac{rt}{R}\partial_r.
\eeq
If $R$ is taken small enough such that $\vev{T_{00}}$ is approximately constant throughout
the ball, equation (\ref{eqn:dSIR}) becomes
\beq\label{eqn:dSIRT}
\delta S_\text{IR} = 2\pi \frac{\Omega_{d-2}R^d}{d^2-1} \delta \vev{T_{00}}. 
\eeq

The assumption of vacuum equilibrium states that $\delta S_\text{tot} = \delta S_\text{UV}
+\delta S_\text{IR} = 0$, and this requirement, along with the expressions (\ref{eqn:dSUVG}) 
and (\ref{eqn:dSIRT}), leads to the relation
\beq
G_{00}+\Lambda g_{00} = \frac{2\pi}{\eta}\delta\vev{ T_{00}},
\eeq
which is recognizable as a component of the Einstein equation with $G_N = \frac1{4\eta}$.  
Requiring that this hold for 
all Lorentz frames and at each spacetime 
point leads to the full tensorial equation, and conservation of $T_{ab}$
and the Bianchi identity imply that $\Lambda(x)$ is a constant.  

The expression of $\delta S_\text{IR}$ in (\ref{eqn:dSIRT}) is  
special to a CFT, and cannot be expected to hold for
more general field theories.  However, it is enough if, in the small $R$ limit, it takes the 
following form
\beq\label{eqn:dSIRmod}
\delta S_\text{IR} = 2\pi \frac{\Omega_{d-2} R^d}{d^2-1} \left(\delta\vev{T_{00}} + C g_{00} \right).
\eeq
Here, $C$ is some scalar function of spacetime, formed from expectation values of operators
in the quantum theory.  With this form of $\delta S_\text{IR}$, 
the requirement that $\delta S_\text{tot}$
vanish in all Lorentz frames and at all points now leads to the tensor equation
\beq\label{eqn:EEtensor}
G_{ab}+\Lambda g_{ab} = \frac{2\pi}{\eta}\left(\delta\vev{T_{ab}}+ C g_{ab}\right).
\eeq
Stress tensor conservation and the Bianchi identity now impose that $\frac{2\pi}{\eta}
C(x) = \Lambda(x)+
\Lambda_0$, and once again the Einstein equation with a cosmological constant is recovered. 

The purpose of the present paper is to evaluate $\delta S_{\text{IR}}$ appearing in equation
(\ref{eqn:dSIRmod}) in a CFT deformed by a relevant operator of dimension $\D$.  
It is crucial in the above derivation that $C$ transform as a scalar under 
a change of Lorentz frame.  As long as this requirement is met, complicated dependence on 
the state or operators in the theory is allowed.  
In the simplest case, $C$ would be given by the variation of some 
scalar operator expectation value, $C = \delta\vev{X}$, with $X$ independent of the quantum 
state, since such an object has trivial transformation properties under Lorentz boosts.  
We find this to be the case for the first order state variations we considered; however, 
the operator $X$ has the peculiar feature that it depends explicitly on the radius of the ball.  
The constant $C$ is found to have a term scaling with the ball size as $R^{2\D-d}$
(or $\log R$ when $\D=\dt$), and 
when $\D\leq\dt$, this term dominates over the stress tensor term as $R\rightarrow 0$.  
Furthermore, as pointed out in \cite{Casini2016a}, even in the CFT where the first order
variation of the entanglement entropy vanishes, the second order piece contains the 
same type of term scaling as $R^{2\D-d}$, which again dominates for small $R$.  
This leads to the conclusion that the local curvature scale $\Lambda(x)$ must be 
allowed to depend on $R$.  
This proposed resolution will be discussed further in section 
\ref{sec:EEimps}.

\subsection{Entanglement entropy of balls in conformal perturbation theory} \label{sec:EEballs} 
Checking the conjecture (\ref{eqn:dSIRmod}) requires  a method for calculating the 
entanglement entropy of balls in a non-conformal theory.  Faulkner has recently 
shown how to perform this calculation in a CFT deformed by a relevant operator, 
$\int f(x)\op(x)$ \cite{Faulkner2015}.  This deformation may be split into two parts, 
$f(x) = g(x) + \lambda(x)$, where the coupling $g(x)$ represents the deformation of the 
theory away from a CFT, while the function $\lambda(x)$ produces a variation of the 
state away from vacuum.  The change in entanglement relative to the CFT vacuum will then
organize into a double expansion in $g$ and $\lambda$,
\beq
\delta S = S_g + S_\lambda+ S_{g^2} + S_{g\lambda} + S_{\lambda^2} +\ldots.
\eeq
The terms in this expansion that are $O(\lambda^1)$ and any order in $g$ are the ones relevant
for $\delta S_\text{IR}$ in equation (\ref{eqn:dSIRmod}).  Terms that are $O(\lambda^0)$ are part
of the vacuum entanglement entropy of the deformed theory, and hence are not of interest 
for the present analysis.  Higher order in $\lambda$ terms may also be relevant,  especially
in the case that
the $O(\lambda^1)$ piece vanishes, which occurs, for example, in a CFT.

We begin with the Euclidean path integral 
representations of the reduced density matrices in the ball $\Sigma$
for the CFT vacuum $\rho_0$ and for
the deformed theory excited state $\rho = \rho_0+\delta\rho$.  The matrix
elements of the vacuum density
matrix are
\beq \label{eqn:rho0pi}
\langle\phi_-| \rho_0 |\phi_+ \rangle = \frac1Z \int_{\substack{\phi(\Sigma_+) = \phi_+\\
\phi(\Sigma_-) = \phi_-}}\pd \phi\, e^{-I_0 }.
\eeq
Here, the integral is over all fields satisfying the boundary conditions  $\phi = \phi_+$
on one side of the surface $\Sigma$, and $\phi=\phi_-$ on the other side.  
The partition function $Z$ 
is represented by an unconstrained path integral,
\beq
Z= \int \pd\phi\, e^{-I_0}.
\eeq
It is useful to think of the path integral (\ref{eqn:rho0pi}) as evolution along an angular
variable $\theta$ from the $\Sigma_+$ 
surface at $\theta=0$ to the $\Sigma_-$ surface at $\theta=2\pi$
\cite{Kabat:1994vj, Holzhey:1994we, Wong2013}.
When this evolution follows the flow of the conformal Killing
vector (\ref{eqn:CKV}) (analytically continued to Euclidean space), it is generated by 
the conserved Hamiltonian $K$ from equation (\ref{eqn:K}).  This leads  to the operator
expression for $\rho_0$ given in equation (\ref{eqn:rho0}).

The path integral representation for $\rho$ is given in a similar manner, 
\begin{align}
\label{eqn:rhopi}
\langle\phi_-| \rho |\phi_+ \rangle &= \frac1N \int_{\substack{\phi(\Sigma_+) = \phi_+\\
\phi(\Sigma_-) = \phi_-}}\pd \phi\, e^{-I_0 -\int f\op}  \\
\label{eqn:rhoexpand}
&=\frac1{Z+\delta Z} \int_{\substack{\phi(\Sigma_+) = \phi_+\\
\phi(\Sigma_-) = \phi_-}}\pd \phi\, e^{-I_0}\left(1-\int f\op +\frac12\iint f\op f\op  -\ldots\right)
\end{align}
Again viewing this path integral as an evolution from $\Sigma_+$ to $\Sigma_-$, with 
evolution operator
$\rho_0 = e^{-2\pi K}/Z$, we can extract the operator expression of $\delta \rho = \rho-\rho_0$, 
\beq\label{eqn:drho}
\delta\rho = -\rho_0\int  f\op +\frac12\rho_0\iint  T\left\{f\op f\op \right\} - \ldots -\text{traces},
\eeq
where $T\{\}$ denotes angular ordering in $\theta$. 
The ``-traces'' terms in this expression arise from  $\delta Z$ in (\ref{eqn:rhoexpand}).  
These terms ensure that $\rho$ is normalized, or equivalently
\beq\label{eqn:traceless}
\tr(\delta\rho) = 0.
\eeq
We suppress writing these terms explicitly since they will play no role in the remainder of this
work.  

Using these expressions for $\rho_0$ and $\delta\rho$, we can now develop the perturbative
expansion of the entanglement entropy, 
\beq
S = -\tr\rho \log\rho .
\eeq
It is useful when expanding out the logarithm to write this 
in terms of the resolvent integral,\footnote{One can also expand the logarithm
using the Baker-Campbell-Hausdorff formula, see e.g.\ \cite{Kelly:2015mna}.}
\begin{align}
S &= \int_0^\infty 
d\beta\left[\tr\left(\frac{\rho}{\rho+\beta}\right) - \frac1{1+\beta}\right]\\
\label{eqn:drhobetaint}
&= S_0 +\tr\int_0^\infty d\beta \frac{\beta}{\rho_0+\beta}\left[\delta\rho\frac{1}{\rho_0+\beta}
-\delta\rho\frac{1}{\rho_0+\beta}\delta\rho\frac{1}{\rho_0+\beta}+\ldots\right].
\end{align}
The first order term in $\delta \rho$ is straightforward to evaluate.  Using the 
cyclicity of the trace and 
equation (\ref{eqn:traceless}), the $\beta$ integral is readily evaluated, and applying 
(\ref{eqn:rho0}) one finds
\beq \label{eqn:dS1}
\delta S^{(1)} = 2\pi \tr(\delta \rho\, K) = 2\pi\delta\vev{K}.
\eeq
Note when $\delta\rho$ is a first order variation, this is simply the first law of entanglement
entropy \cite{Blanco:2013joa} (see also \cite{Bhattacharya:2012mi}). 

\begin{figure}
\centering
\includegraphics[width=0.65\textwidth]{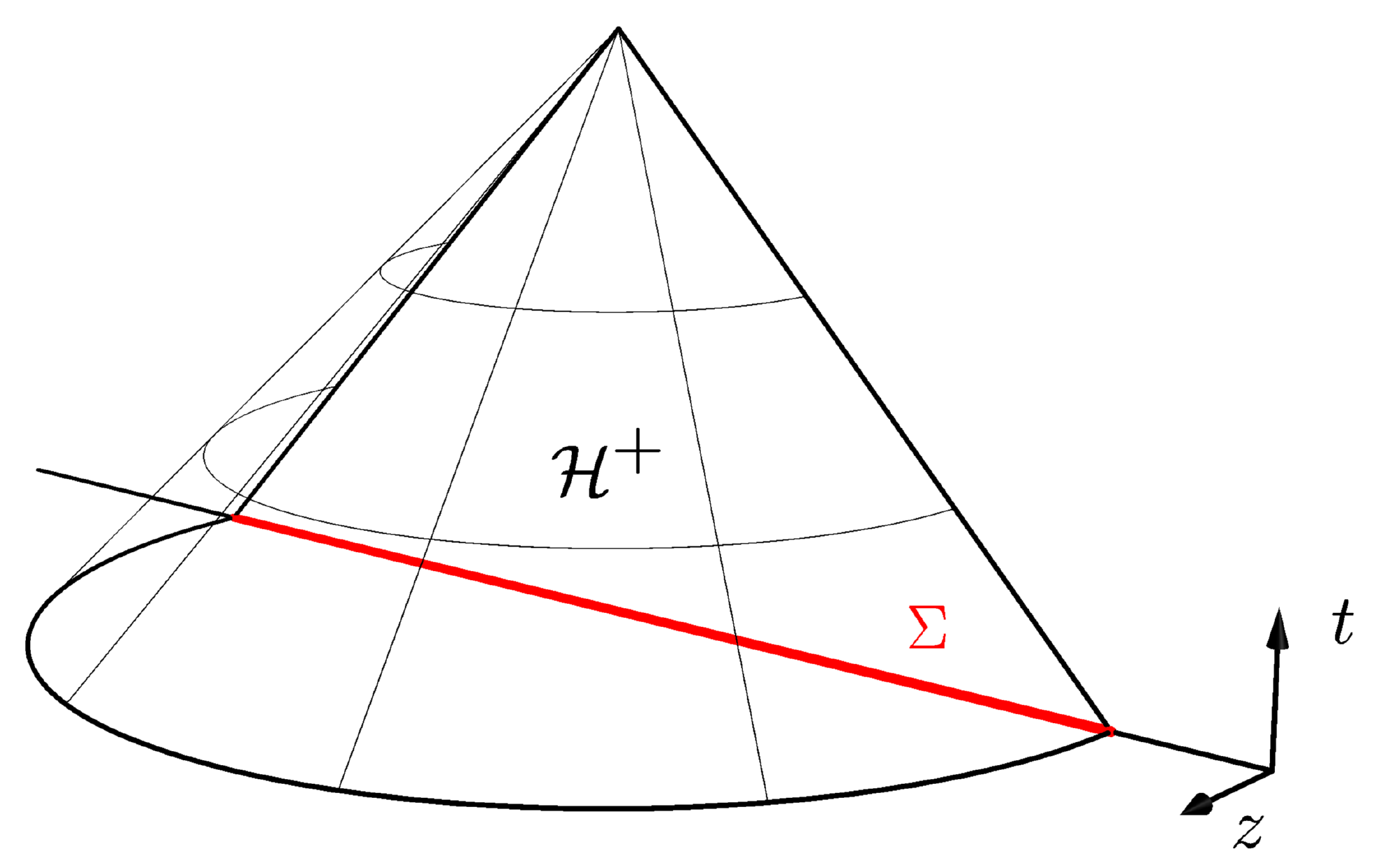}
\caption{Bulk AdS-Rindler horizon $\mathcal{H}^+$.  The horizon extends from the 
bifurcation surface in the bulk at $t=0$ along the cone to the tip at $z=0$, $t=R$.  
The ball-shaped surface $\Sigma$ in the boundary CFT shares a boundary
with the bifurcation surface at $t=z=0$. \label{fig:AdSRindler}}

\end{figure}

The second order piece of (\ref{eqn:drhobetaint}) is more involved, and much of reference
\cite{Faulkner2015} is devoted to evaluating this term.  
The surprising result is that this term may
be written holographically as the 
flux through an emergent  AdS-Rindler horizon of a conserved energy-momentum
current for a scalar field\footnote{Reference 
\cite{Faulkner2015} further showed that this is equivalent
to the Ryu-Takayanagi prescription for calculating the entanglement entropy
\cite{Ryu:2006bv, Ryu:2006ef}, using
an argument similar to the one employed in \cite{Faulkner:2013ica} 
deriving the bulk linearized Einstein equation from
the Ryu-Takayanagi formula.}  
(see figure \ref{fig:AdSRindler}).
The bulk scalar field $\phi$ satisfies the free Klein-Gordon equation in AdS with mass
$m^2 = \D(\D-d)$, as is familiar from the usual holographic dictionary \cite{Witten:1998qj}. 
The  specific AdS-Rindler horizon that is used is the one with
a bifurcation surface that asymptotes near the boundary 
to the entangling surface $\partial\Sigma$
in the CFT.
This result holds for {\it any} CFT, including those which are not normally considered 
holographic. 

We now describe the bulk calculation in more detail.  Poincar\'{e} coordinates are used
in the bulk, where the metric takes the form
\beq
ds^2 = \frac1{z^2} \left( -dt^2+dz^2+dr^2 + r^2 d\Omega_{d-2}^2\right).
\eeq
The coordinates $(t,r,\Omega_i)$ match onto the Minkowski coordinates of the CFT at the 
conformal boundary $z=0$.  The conformal Killing vector $\zeta^a$ of the CFT, defined in
equation (\ref{eqn:CKV}), extends to a Killing vector in the bulk,
\beq
\xi = \left(\frac{R^2-t^2-z^2-r^2}{2R}\right)\partial_t - \frac{t}{R}(z\partial_z + r\partial_r).
\eeq
The Killing horizon $\mathcal{H^+}$ 
of $\xi^a$ defines the inner boundary of the AdS-Rindler patch for $t>0$, and sits at
\beq
r^2+z^2 = (R-t)^2.
\eeq

The contribution of the second order piece of (\ref{eqn:drhobetaint}) to the entanglement entropy
is 
\beq \label{eqn:H+int}
\delta S^{(2)} = -2 \pi\int_\mathcal{H^+} d\Sigma^a \xi^b T^B_{ab},
\eeq
where the integral is over the horizon to the future of the bifurcation surface at $t=0$. 
The surface element on the horizon is $d\Sigma^a = \xi^a d\chi dS$, where $\chi$ is a
parameter for $\xi^a$ satisfying $\xi^a\nabla_a\chi=1$,
and $dS$ is the area element in the transverse space.  $T^B_{ab}$
is the stress tensor of a scalar field $\phi$ satisfying the Klein-Gordon equation,
\beq \label{eqn:bulkeom}
\nabla_c\nabla^c \phi -\D(\D-d) \phi = 0.
\eeq
Explicitly, the stress tensor is
\beq
T^B_{ab} = \nabla_a\phi\nabla_b\phi-\frac12(\D(\D-d)\phi^2+\nabla_c\phi \nabla^c\phi) g_{ab},
\eeq
which may be rewritten when $\phi$ satisfies the field equation (\ref{eqn:bulkeom}) as 
\beq\label{eqn:TBab}
T^B_{ab} = \nabla_a\phi \nabla_b\phi - \frac14 g_{ab}\nabla_c\nabla^c \phi^2.
\eeq

The boundary conditions for $\phi$ come about
from its defining integral,
\beq \label{eqn:phiint}
\phi(x_B) = \frac{\Gamma(\D)}{\pi^{\dt} \Gamma(\D-\dt)} 
\int_{C(\delta)} d \tau \int d^{d-1} \vec{x} \frac{z^\D f(\tau,\vec{x})}{
\left(
z^2+(\tau-i t_B)^2+(\vec{x}-\vec{x}_B)^2\right)^\D},
\eeq
where $x_B = (t_B, z, \vec{x}_B)$ are the real-time bulk coordinates, and $(\tau,\vec{x})$ are
coordinates on the boundary Euclidean section.  The normalization of this field arises from
a particular choice of the normalization for the $\op\op$ two-point function,
\beq \label{eqn:cD}
\vev{\op(x)\op(0)} = \frac{c_\D}{x^{2\D}},\qquad c_\D = \frac{(2\D-d)\Gamma(\D)}{\pi^{\dt}\Gamma
(\D-\dt)},
\eeq
which is chosen so that the relationship (\ref{eqn:phibdyEuc}) holds.  Note that 
sending $c_\D\rightarrow \alpha^2 c_\D$ multiplies $\phi$ by a single factor of $\alpha$.  
The integrand in (\ref{eqn:phiint}) has branch points at $\tau = i\left(t_B\pm\sqrt{z^2+
(\vec{x}-\vec{x}_B)^2}\right)$, and the branch cuts extend along the imaginary axis to 
$\pm i\infty$.  The notation $C(\delta)$ on the $\tau$ integral refers to the $\tau$ contour
prescription, which must lie along the real axis and be cut off near $0$ at $\tau=\pm\delta$.  
This can lead to a divergence in $\delta$ when the contour is close to the branch point (which
can occur when $t_B\sim \sqrt{z^2+(\vec{x}-\vec{x}_b)^2}$), and this ultimately cancels
against a divergence in  $\vev{T_{00} \op \op}$ from $\delta S^{(1)}$.  More details about 
these divergences and the origin of this contour and branch 
prescription can be found in \cite{Faulkner2015}.

From equation (\ref{eqn:phiint}), one can now read off the boundary conditions
as $z\rightarrow 0$.  The solution
should be regular in the bulk, growing at most like $z^{d-\D}$ for large $z$ if $f(\tau,\vec{x})$ is 
bounded.  On the Euclidean section $t_B=0$, it behaves for $z\rightarrow 0$ as
\beq \label{eqn:phibdyEuc}
\phi\rightarrow f(0,\vec{x_B}) z^{d-\D} +\beta(0,\vec{x_B}) z^\D,
\eeq
where the function $\beta$ may be determined by the integeral (\ref{eqn:phiint}), but also 
may be fixed by demanding regularity of the solution in the bulk. This is consistent with 
the usual holographic dictionary \cite{Klebanov1999, Balasubramanian:1998de}, 
where $f$ corresponds to the coupling, and $\beta$ is related
to $\vev{\op}$ by\footnote{The minus sign appearing here is due to the source in the 
generating functional being $-\int f\op$ as opposed to $\int f\op$}
\beq\label{eqn:beta}
\beta(x) = \frac{-1}{2\D-d} \vev{\op(x)}.
\eeq
This formula follows from defining the renormalized expectation value $\vev{\op}$ using 
a holographically renormalized two-point function,
\beq
\mvev{\big}{\op(0)\op(x)}^{z,\text{ren.}} 
= \frac{c_\D}{(z^2 + x^2)^\D} - (2\D-d)z^{d-2\D}  \delta^d(x).
\eeq  
The $\delta$ function in this formula subtracts off the divergence near
$x= 0$.  Using the renormalized two-point function, the expectation value of $\op$ at 
first order in $f$ is
\beq
\vev{\op(x)} = -\int d^d y f(y) \mvev{\Big}{\op(x) \op(y)}^{z,\text{ren.}},
\eeq  
and by comparing this formula to (\ref{eqn:phiint}) at small values $z$ and $t_B=0$, 
one arrives at 
equation (\ref{eqn:beta}).

In real times beyond $t_B >z$, $\phi(x_B)$ has only a $z^\D$ component near $z=0$.  The
integral effectively shuts off the coupling $f$ in real times.  This follows from the 
use of a Euclidean path integral to define the state; other real-time behavior may be 
achievable using the Schwinger-Keldysh formalism.
When $t_B\sim z$, there are divergences associated with switching off the coupling in real
times, and these are regulated with the $C(\delta)$ contour prescription.

\begin{figure}
\centering
\includegraphics[width=0.4\textwidth]{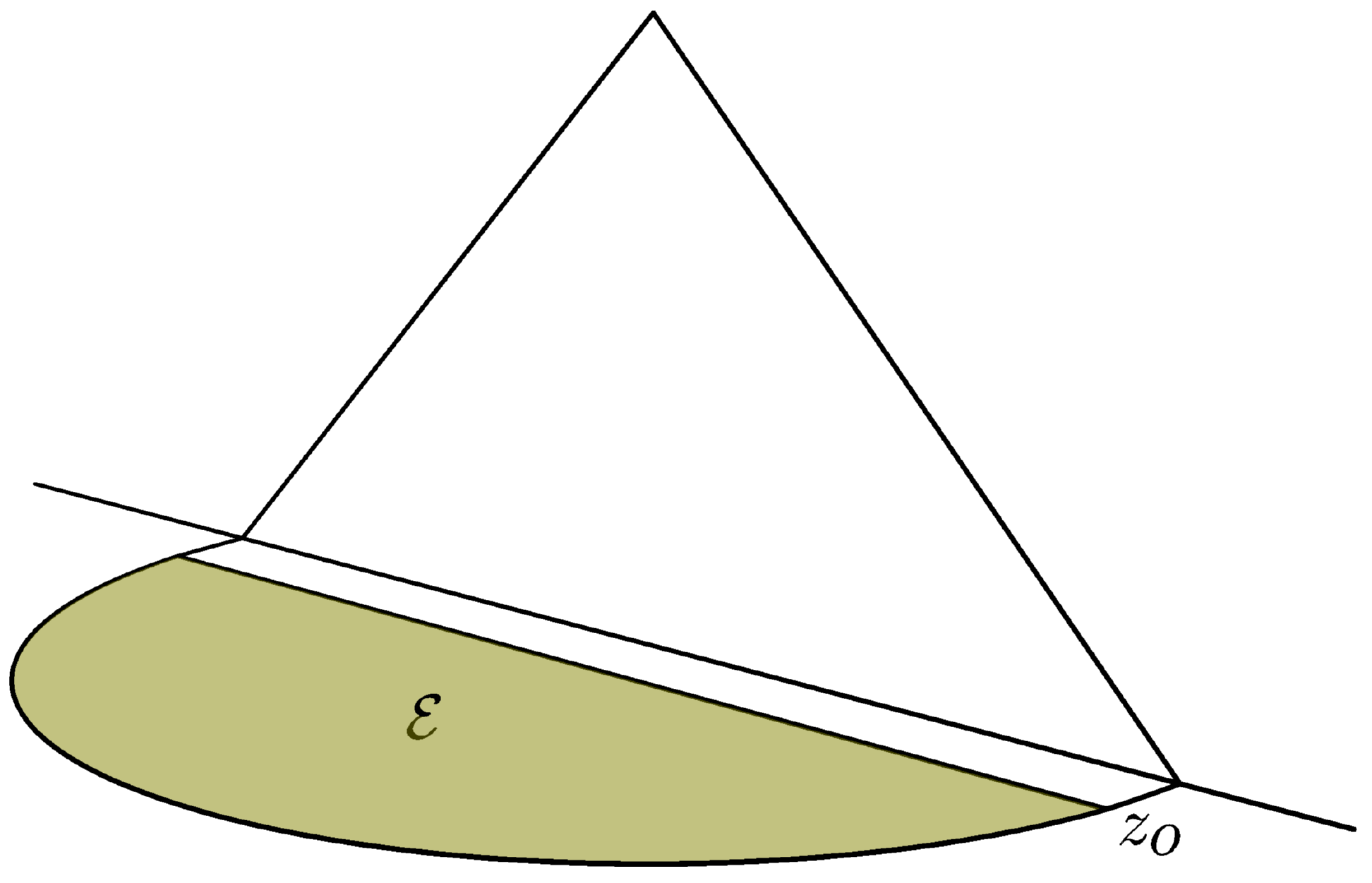}\qquad
\includegraphics[width=0.4\textwidth]{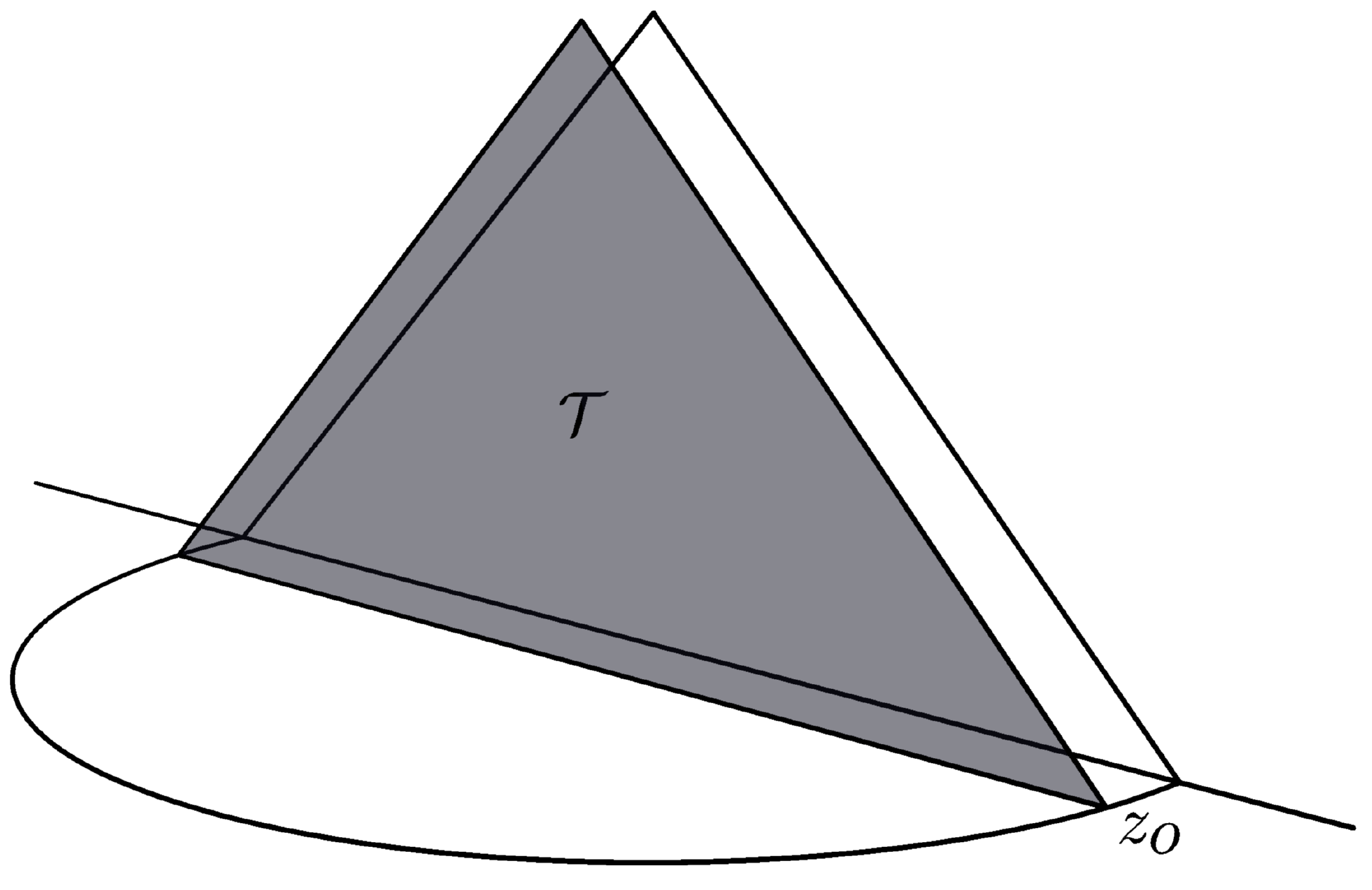}
\caption{$\mathcal{E}$ and $\mathcal{T}$ surfaces over which the flux integrals (\ref{eqn:Eint})
and (\ref{eqn:Tint}) are computed. \label{fig:ET}}

\end{figure}

Returning to the flux equation (\ref{eqn:H+int}), 
since $\xi^a$ is a Killing vector, this integral defines a conserved quantity, and 
may be evaluated on any other surface homologous to 
$\mathcal{H}^+$.  The choice which is
most tractable is to push the surface down to $t_B=0$, where the Euclidean AdS solution can 
be used to evaluate the stress tensor.  The $t_B=0$ surface $\mathcal{E}$ 
covers the region between the 
horizon and $z=z_0$, where it must be cut off to avoid a divergence in the integral.  To remain
homologous to $\mathcal{H^+}$, this must be supplemented by a timelike surface $\mathcal{T}$ 
at the 
cutoff $z=z_0$ which extends upward to connect back with $\mathcal{H}^+$.  In the limit
$z_0\rightarrow 0$, the surface $\mathcal{T}$ approaches the domain of dependence
$D^+(\Sigma)$ of the ball-shaped region in the CFT (see figure \ref{fig:ET}).  Finally, there
will  be a contribution from a 
region along the original surface $\mathcal{H}^+$ between $z_0$ and $0$, 
but in the limit $z_0\rightarrow 0$, the contribution to the integral from this surface will  
vanish.\footnote{
This piece may become important in the limiting case $\D=\dt-1$, which requires special 
attention.  We will not consider this possibility further here.}

Using equation (\ref{eqn:TBab}), the integral on the surface $\mathcal{E}$ 
can be written out more explicitly: 
\begin{align}
&-2\pi\int_\mathcal{E}d\Sigma^a\xi^b T^B_{ab} \nonumber \\
\label{eqn:Eint}
&= 2\pi\int d\Omega_{d-2}\int_{z_0}^R 
\frac{dz}{z^{d-1}} \int_0^{\sqrt{R^2-z^2}} dr\, r^{d-2} \left[\frac{R^2-r^2-z^2}{2R}\right]\left[
(\partial_\tau \phi)^2-\frac{ \nabla^2_E\phi^2}{4z^2}\right].
\end{align}
This formula uses the solution on the Euclidean section in the bulk, with Euclidean 
time $\tau_B = i t_B$.  This is acceptable on the $t_B=0$ surface since the stress tensor
there satisfies $T_{\tau\tau}^B = -T^B_{tt}$.  The Laplacian $\nabla_E^2$ is hence the 
Euclidean AdS Laplacian.
The  
$\mathcal{T}$ surface integral
is 
\begin{align}
&2\pi \int_{\mathcal{T}} d\Sigma^a \xi^bT^B_{ab} \nonumber \\
\label{eqn:Tint}
&= \frac{2\pi}{z_0^{d-1}} \int d\Omega_{d-2} \int_0^R dt\int_0^{R-t} dr\, r^{d-2}\left\{
\left[\frac{R^2-r^2-t^2}{2R}\right]\partial_z\phi\partial_t\phi - \frac{z_0t}{R} \left[
(\partial_z\phi)^2-\frac{ \nabla^2\phi^2 }{4z_0^2}\right]
\right\}. 
\end{align}
Here, note that the limits of integration have been set to coincide with $D^+(\Sigma)$, which 
is acceptable when taking $z_0\rightarrow0$.

\section{Producing excited states} \label{sec:excited}
This section describes the class of states that are formed from the Euclidean path integral
prescription, and also discusses restrictions on the source function $f(x)$.  One requirement
is that the density matrix be Herimitian.  For a density matrix constructed from a path integral
as in (\ref{eqn:rhopi}), this translates to the condition that the deformed action $I_0 + \int f\op$
be reflection symmetric about the $\tau=0$  surface on which the state is evaluated.  When
this is satisfied, $\rho$ defines a pure state \cite{Cooperman:2013iqr}.  
Since this imposes  $f(\tau, \vec{x}) = f(-\tau,\vec{x})$, it gives
the useful condition
\beq
\partial_\tau f(0,\vec{x})=0,
\eeq
which simplifies the evaluation of the bulk integral (\ref{eqn:Eint}).  

Another condition on the state is that the stress tensor $T^g_{ab}$ of the deformed theory
and the operator $\op$ have non-divergent expectation values, compared to the vacuum.  
These divergences are not independent, but are related to each other through Ward identities.  
The $\vev{\op}$ divergence is straightforward to evaluate, 
\begin{align}
\vev{\op(0)} &= \frac{1}{N}\int \pd\phi e^{-I_0}\left(1-\int f\op+\ldots\right)\op(0) \\
&= -\int_{C(\delta)} d^dx f(x)\mvev{\Big}{\op(0)\op(x)}_0 ,
\end{align}
where the $0$ subscript indicates a CFT vacuum correlation function.  $C(\delta)$
refers to the regularization of this correlation function, which  is a  point-splitting
cutoff for  $|\tau| <\delta$.   Note that $\delta$ is the same regulator appearing in 
the 
definition of the bulk scalar field, equation (\ref{eqn:phiint}).  

Only the change  $\delta\vev{\op}$  in this correlation
function
relative to the deformed theory vacuum must be free of divergences. From the 
decomposition $f(x) = g(x)+\lambda(x)$, with $g(x)$ representing the 
deformation of the theory and $\lambda(x)$ the state deformation, one finds that 
the divergence in $\delta\vev{\op}$ comes from the coincident limit $x\rightarrow 0$.  It can
be extracted by expanding $\lambda(x)$ around $x=0$.  The leading divergence is then
\begin{align}
\delta\vev{\op(0)}_\text{div} &= -\lambda(0)\int_{C(\delta)}d\tau \int d\Omega_{d-2}\int_0^\infty dr 
\frac{r^{d-2}c_\D}{
(\tau^2 +r^2)^\D} \nonumber\\
&= 
\label{eqn:dOdiv}
-\lambda(0)\frac{2\Gamma(\D-\dt+\frac12)}{\sqrt{\pi} \,\Gamma(\D-\dt)} 
\delta^{d-2\D}
\end{align}
When $\D\geq\dt$, a divergence in $\delta\vev{\op}$ exists unless $\lambda(0)=0$.\footnote{When
$\D=\dt$, after appropriately redefining $c_\D$ (see equation (\ref{eqn:cD'})), 
it becomes a $\log\delta$ divergence.}  
Further, this must hold at every point on the 
$\tau=0$ surface, which leads to the requirement that $\lambda(0,\vec{x})=0$.  Additionally, there
can be subleading divergences proportional to $\delta^{d-2\D+2n} \partial_\tau^{2n}\lambda(0,
\vec{x})$ for all integers $n$ where the $\delta$ 
exponent is negative or zero.\footnote{Divergences
proportional to the spatial derivative of $\lambda$ are not present since the condition from the
 leading divergence
already set these to zero.}
Thus, the requirement on $\lambda$ is that its first $2q$ $\tau$ derivatives should vanish at
$\tau=0$, where 
\beq\label{eqn:q}
q=\left \lfloor \D-\dt\right\rfloor.
\eeq

We can also check that this condition leads to a finite value expectation value for the stress 
tensor, which for the deformed theory is 
\beq \label{eqn:Tgabdef}
T^g_{ab} = \frac{2}{\sqrt{g}} \frac{\delta I}{\delta g^{ab}} = T^0_{ab}-g\op g_{ab},
\eeq
where $T^0_{ab}$ is the stress tensor for the CFT.  For the $T^0_{\tau\tau}$ component,  the 
expectation value is 
\begin{align}
\vev{T^0_{\tau\tau}(0)} 
&=\frac12\iint_{C(\delta)} d^d x \,d^d y f(x) f(y)\mvev{\Big}{T^0_{\tau\tau}(0)\op(x)\op(y)}_0.
\end{align}
The divergence in this correlation function comes from $x,y\rightarrow 0$ simultaneously.
It can be evaluated by expanding $f$ around $0$, and then employing Ward identities to 
relate it to the $\op\op$ two-point function (see, e.g.\  section \ref{sec:D=d2calc} of this 
paper or 
Appendix D of \cite{Faulkner2015}).  
The first order in $\lambda$ piece, which gives $\delta\vev{T^0_{\tau\tau}}$, is
\beq
\delta\vev{T^0_{\tau\tau}}_\text{div}
 = -g \lambda(0) 2^{d-2\D}\frac{2\Gamma(\D-\dt+\frac12)}{\sqrt{\pi}\,
\Gamma(\D-\dt)} \delta^{d-2\D}.
\eeq
The divergence in the actual energy density also receives a contribution from the $\op$ 
divergence (\ref{eqn:dOdiv}).  Using (\ref{eqn:Tgabdef}), this is found to be
\beq
\delta\vev{T^g_{\tau\tau}}_\text{div} = -g\lambda(0)\frac{2\Gamma(\D-\dt+\frac12)}{\sqrt{\pi}\,
\Gamma(\D-\dt)} (2^{d-2\D}-1) \delta^{d-2\D}.
\eeq
As with the $\delta\vev{\op}$ divergence,  requiring that $\lambda(0,\vec{x})=0$ ensures that 
the excited state has finite energy density.\footnote{Curiously, the divergences in $T^g_{ab}$
cancel without imposing $\lambda(0)=0$ when $\D=\dt$. }  
Subleading divergences and other components of 
$T^g_{ab}$ can be evaluated in a 
similar way, and lead to the same requirements on $\lambda$ as were found for the $\op$ 
divergences.

\section{Entanglement entropy calculation} \label{sec:harder} 
Now we compute the change in entanglement entropy for the state formed by the path
integral with the deformed action $I=I_0+\int f\op$, with $f(x) = g(x)+\lambda(x)$ being
a sum of the theory deformation $g$ and the state deformation $\lambda$.
The bulk term $\delta S^{(2)}$ in 
plays an important role in this case.\footnote{A slightly simpler situation would be to
consider the deformed action $I = I_0 + \int g \op + \int \lambda\op_s$, with $\D\neq\D_s$.
Then $\delta S^{(2)}$ gives no contribution at first order in $\lambda$, since this term
arises from the $\op\op_s$ two point function, which vanishes.  However, in this case, the 
term at second order in $\lambda$ would receive a contribution from $\delta S^{(2)}$, and 
it is computed in precisely the same way as described in this section.  Hence we do not 
focus on this case where $\D\neq\D_s$.}    
To evaluate this term, we need the solution for the scalar field in the bulk subject to the 
boundary conditions described in section \ref{sec:EEballs}.  Since $\phi$ satisfies
a linear field equation, so we may solve separately for the solution corresponding to $g$ and 
the solution corresponding to $\lambda$.  The function $g(x)$ is taken to be spatially constant, 
and either constant in Euclidean time or set to zero at some IR length scale $L$.  Its solution
is most readily found by directly evaluating the integral (\ref{eqn:phiint}), and we will discuss
it separately in each of the cases $\D>\dt$, $\D<\dt$ and $\D=\dt$ considered below.  

The solution for $\lambda(x)$ takes the same form in all three cases, so we begin by describing
it.  On the Euclidean section in Poincar\'{e} coordinates, the field equation (\ref{eqn:bulkeom}) 
is
\beq\label{eqn:eompoin}
\left[ z^{d+1}\partial_z(z^{-d+1}\partial_z)+z^2\left(\partial_\tau^2 +
r^{-d+2}\partial_r(r^{d-2}\partial_r) +r^{-2}\nabla^2_{\Omega_{d-2}} \right)\right]\phi -
\D(\D-d) \phi = 0,
\eeq
where $\nabla^2_{\Omega_{d-2}}$ denotes the Laplacian on the $(d-2)$-sphere.  
Although one may consider arbitrary spatial dependence for the function $\lambda(x)$,
the present calculation is concerned with the small ball limit, where the state may be taken
uniform across the ball.  We therefore restrict to $\lambda = \lambda(\tau)$.  One can 
straightforwardly generalize to include corrections due to spatial dependence in $\lambda$,
and these will produce terms suppressed in powers of $R^2$.  

Equation (\ref{eqn:eompoin}) 
may be solved by separation of variables.  The $\tau$ dependence is given by
$\cos(\omega\tau)$, since it must be $\tau$-reflection symmetric.  
This leads to the equation for the $z$-dependence,  
\beq
\partial_z^2\phi-\frac{d-1}{z}\partial_z\phi -\left(\omega^2+\frac{\D(\D-d)}{z^2}\right)\phi = 0.
\eeq
This has modified Bessel functions as solutions, and regularity as $z\rightarrow\infty$ selects 
the solution proportional to $z^{\dt} K_\alpha(\omega z)$, with
\beq
\alpha = \dt-\D.
\eeq
Hence, the final  bulk solution is
\beq \label{eqn:phio}
\phi_\omega = \lambda_{\omega}\left(\frac{\omega}{2}\right)^{\D-\dt}
\frac{2z^{\dt}K_\alpha(\omega z)}{\Gamma(\D-\dt)} \, \cos\omega \tau.
\eeq
where the normalization has been chosen so that the coefficient of $z^{d-\D}$
in the near-boundary expansion is
\beq \label{eqn:lambda}
\lambda = \lambda_\omega \cos(\omega\tau).
\eeq
A single frequency solution will not satisfy the requirement derived in section \ref{sec:excited}
that $\lambda(0,\vec{x})$ and its first $2q$ $\tau$-derivatives vanish (where $q$ was
given in 
(\ref{eqn:q})).  
Instead, $\lambda$ must be constructed from a wavepacket of several frequencies,
\beq
\lambda(\tau) = \int_0^\infty d\omega \lambda_\omega \cos(\omega\tau),
\eeq
with Fourier components $\lambda_\omega$ satisfying
\beq \label{eqn:lambdaoint}
\int_0^\infty d\omega\, \omega^{2n} \lambda_\omega = 0
\eeq
for all nonnegative integers $n\leq q$.  Finally, the coefficients $\lambda_\omega$ should 
fall off rapidly before $\omega$ becomes larger than $R^{-1}$, since such a state would be 
considered highly excited relative to the scale set by the ball size.

Using these solutions, we may proceed with the entanglement entropy calculation.  The answer
for $\D>\dt$ in section \ref{sec:D>dt}
comes from a simple application of the formula derived in \cite{Faulkner2015}.  
In section \ref{sec:D<dt} when considering 
$\D<\dt$, we must introduce a new element into the calculation to deal with IR divergences
that arise.  This is just a simple IR cutoff in the theory deformation $g(x)$, 
which allows a finite answer to emerge, although a new set of divergences along the timelike
surface $\mathcal{T}$ must be shown to cancel. A similar story emerges in section \ref{sec:D=dt}
for $\D=\dt$,
although extra care must be taken due to the presence of logarithms in the solutions.   

\subsection{$\D >\dt$} \label{sec:D>dt}
The full bulk scalar field separates into two parts, 
\beq
\phi = \phi_0 + \phi_\omega,
\eeq
with $\phi_\omega$ from (\ref{eqn:phio}) describing the state deformation, while $\phi_0$ 
corresponds to the theory deformation $g(x)$. Since no IR divergences arise at this order in
perturbation theory when $\D>\dt$, we can take $g$ to be constant everywhere.  The solution
in the bulk on the Euclidean section then takes the simple form
\beq
\phi_0 = g z^{d-\D}.
\eeq

Given these two solutions, the bulk contribution to $\delta S^{(2)}$ may be computed using 
equation (\ref{eqn:Eint}).  Note that $\partial_\tau\phi=0$ on the $\tau=0$ surface, so we 
only need the $\nabla^2 \phi^2$ term in the integrand.  Before evaluating this term, it is useful
to expand $\phi_\omega$ near $z=0$.  This expansion takes the form
\beq\label{eqn:phioseries}
\phi_\omega = \left[\lambda_\omega z^{d-\D}\sum_{n=0}^{\infty} a_n (\omega z)^{2n}
+ \beta_\omega z^{\D}\sum_{n=0}^\infty b_n (\omega z)^{2n}\right]\cos(\omega\tau),
\eeq
where 
\beq\label{eqn:bo}
\beta_\omega = \lambda_\omega \left(\frac\omega2\right)^{2\D-d} \frac{\Gamma(\dt-\D)}{\Gamma(
\D-\dt)},
\eeq
and the coefficients $a_n$ and $b_n$ are given in appendix \ref{sec:Bessel}.  The $O(\lambda^1)$
term in $\phi^2$ is $2\phi_0\phi_\omega$, and this modifies the power series
(\ref{eqn:phioseries}) by changing the leading powers to $z^{2(d-\D)}$ and $z^d$.  The 
Laplacian in the bulk is 
\beq
\nabla^2 = z^2\partial_\tau^2+z^{d+1}\partial_z(z^{-d+1}\partial_z).
\eeq
Acting on the $\phi_0\phi_\omega$ series, the effect of the $\tau$ derivative is to multiply
by $-\omega^2z^2$, which shifts each term to one higher term in the series.  The $z$ derivatives
do no change the power of $z$, but rather multiply each term by a constant, 
$2(d-\D+n)(d-2\D+2n)$
for the $a_n$ series and $2  n(d+2n)$ for the $b_n$ series (note in particular it annihilates the 
first term in the $b_n$ series).  After this is done, the series may be reorganized for $\tau=0$ as
\beq \label{eqn:cndn}
2\nabla^2
\phi_0\phi_\omega = 2g\lambda_\omega z^{2(d-\D)}\sum_{n=0}^\infty c_n(\omega z)^{2n} +
2g\beta_\omega z^d\sum_{n=1}^\infty d_n(\omega z)^{2n},
\eeq
with the coefficients $c_n$ and $d_n$ computed in appendix \ref{sec:Bessel}.

From this, we simply need to evaluate the integral (\ref{eqn:Eint}) for each term in the series.
For a given term of the form $A z^\eta$, the contribution to $\delta S^{(2)}$ is 
\begin{align}
\delta S^{(2)}_\eta &
= -\frac\pi2 \Omega_{d-2} \int_{z_0}^R \frac{dz}{z^{d+1}} \int_0^{\sqrt{R^2-z^2}}
dr\, r^{d-2} \left[\frac{R^2-r^2-z^2}{2R}\right] A z^\eta \\
\label{eqn:Eresult}
&=-A\frac{\pi\Omega_{d-2}}{4(d^2-1)} \left[R^\eta\,\frac{\Gamma(\dt+\frac32)\Gamma(\frac\eta2
-\dt)}{\Gamma(\frac\eta2+\frac32)}  + \frac{R^d z_0^{\eta-d}
\tensor[_2]{F}{_1}\left(-\frac{d+1}{2}, \frac{\eta-d}{2}; \frac{\eta-d}{2}+1;\frac{z_0^2}{R^2} \right)}
{\frac\eta2-\dt}  
 \right].
\end{align}
The second term in this expression contains a set of divergences at $z_0\rightarrow0$ for all
values of $\eta<d$.  These arise exclusively from the $c_n$ series in (\ref{eqn:cndn}). 
In general, the expansion of the hypergeometric function near $z_0=0$ can produce 
subleading divergences, which mix between different terms from the series (\ref{eqn:cndn}).  
These divergences
eventually must cancel against compensating divergences that arise from the $\mathcal{T}$
surface integral in (\ref{eqn:Tint}).  Although we do not undertake a systematic study of 
these divergences, we may assume that they cancel out because the cutoff surface at $z_0$ was 
chosen arbitrarily, and the original integral (\ref{eqn:H+int}) made no reference to it.  Thus,
we may simply discard these $z_0$ dependent divergences, and are left with only the 
first term in   (\ref{eqn:Eresult}).\footnote{When $\eta=d+2j$ for an integer $j$, there are 
subtleties
related to the appearance of $\log z_0$ divergences.  These cases arise when $\D=\dt+m$
with $m$ an integer.  We leave analyzing this case for future work.}  

There is another reason for discarding the $z_0$ divergences immediately: they only arise
in states with divergent energy density.  The coefficient of a term with a $z_0$ divergence
is $2gc_n \omega^{2n} \lambda_\omega$.  The final answer for the entanglement entropy 
will involve integrating over all values of $\omega$. But the requirement of finite energy 
density (\ref{eqn:lambdaoint}) shows that all terms with $n\leq q$, corresponding to
$\eta\leq2d-2\D+2q$, will vanish from the final result.  Given the definition of $q$ in 
(\ref{eqn:q}), these are precisely the terms in (\ref{eqn:Eresult}) that have divergences
in $z_0$.  Note that since $\beta_\omega\propto \omega^{2\D-d}$, which is generically
a non-integer power, the integral over $\omega$ will not vanish, so all the $\beta_\omega$
terms survive.

The resulting bulk contribution to the entanglement entropy at order $\lambda g$ is
\begin{align}
\delta S^{(2)}_{\mathcal{E},\lambda g} = -\frac{g \pi^{\dt+\frac12} }{4} \int_0^\infty d\omega
\left[
\lambda_\omega R^{2(d-\D)} \sum_{n=q+1}^\infty \right. 
& c_n \frac{\Gamma(\dt -\D +n)}{\Gamma(d-\D +\frac32+n)}(\omega R)^{2n}  \nonumber  \\
\label{eqn:dS2Elg}
+\; \beta_\omega R^d \sum_{\hphantom{+} n=1 \hphantom{+} }^\infty 
& \left.   \vphantom{\sum_{n=1}^\infty }
d_n \frac{\Gamma(\dt + n) }{\Gamma(\dt + \frac32+n)} (\omega R)^{2n}
\right].
\end{align}
This expression shows that the lowest order pieces scale as $R^{2(d-\D+q+1)}$ and 
$R^{d+2}$, which both become subleading with respect to the $R^d$ scaling 
of the $\delta S^{(1)}$ piece for small ball size.  Note that a similar technique could 
extend this result to spatially dependent $\lambda(x)$, and simply would amount to an
additional series expansion.

One could perform a similar analysis for the $O(\lambda^2)$ contribution from $\delta S^{(2)}$.  
The series of $\nabla^2\phi_\omega\phi_{\omega'}$ would organize into three series, with 
leading coefficients $\lambda_\omega\lambda_{\omega'} z^{2(d-\D)}$, $(\beta_\omega 
\lambda_{\omega'} +\lambda_\omega\beta_{\omega'})z^d$, and $\beta_\omega \beta_{\omega'}
z^{2\D}$. After integrating over $\omega$ and $\omega'$, and noting which terms vanish 
due to the requirement (\ref{eqn:lambdaoint}), one would find the leading contribution
going as $\beta^2R^{2\D}$.  The precise value of this term is 
\beq\label{eqn:R2Dlambda2}
\delta S^{(2)}_{\lambda^2} = -\frac{\pi\Omega_{d-2}}{d^2-1} R^{2\D}\big(\delta\vev{\op}\big)^2 
\,\frac{\D \Gamma(\dt 
+\frac32) \Gamma(\D-\dt+1)}{(2\D -d)\Gamma(\D+\frac32)},
\eeq
which is quite similar to the $R^{2\D}$ term in equation (\ref{eqn:Dneqdt}).
This is again subleading when $\D>\dt$, but the same terms 
show up for $\D\leq\dt$
in sections \ref{sec:D<dt} and \ref{sec:D=dt}, where they become 
the dominant contribution when $R$ is taken small enough.  The importance of these
second order terms in the small $R$ limit was first noted in \cite{Casini2016a}.

The remaining pieces to calculate come from the integral over $\mathcal{T}$ given by 
(\ref{eqn:Tint}), and $\delta S^{(1)}$ in (\ref{eqn:dS1}), 
which just depends on $\delta\vev{T^0_{00}}$.  When
$\D>\dt$, the only contribution from the $\mathcal{T}$ surface integral is near $t_B\sim z
\rightarrow 0$.  These terms were analyzed in appendix E of \cite{Faulkner2015}, and were
found to give two types of contributions.  The first were counter terms that cancel against
the divergences in the bulk as well as the divergence in $\delta S^{(1)}$.  Although subleading
divergences were not analyzed, these can be expected to cancel in a predictable way.  We
also already argued that such terms are not relevant for the present analysis, due to the 
requirement of finite energy density.  The second type of term is finite, and takes the form
\beq\label{eqn:intDb}
\delta S^{(2)}_{\mathcal{T},\text{finite}} = -2\pi\D\int_{\Sigma} \zeta^t g \beta.
\eeq
The relation between $\beta$ and $\delta\vev{\op}$ identified in (\ref{eqn:beta}) implies
from equation (\ref{eqn:bo}),
\beq \label{eqn:dvevoD>dt}
\delta\vev{\op} = \lambda_\omega\frac{2\Gamma(\dt-\D+1)}{\Gamma(\D-\dt)}
\left(\frac{\omega}{2}\right)^{2\D-d},
\eeq
and assuming the ball is small enough so that this expectation value may be considered 
constant, (\ref{eqn:intDb}) evaluates to
\beq \label{eqn:dS2Tfin}
\delta S^{(2)}_{\mathcal{T},\text{finite}} = 2\pi \frac{\Omega_{d-2} R^d}{d^2-1} \left[\frac{\D}{2\D-d}
g\delta\vev{\op}\right].
\eeq
Similarly, taking $\delta \vev{T^0_{00}}$  to be constant over the ball, the final contribution is
the variation of the modular Hamiltonian piece, given by
\beq\label{eqn:dS1gl}
\delta S^{(1)} = 2\pi \int_\Sigma \zeta^t\delta\vev{T^0_{00}} = 2\pi \frac{\Omega_{d-2} R^d}{d^2-1}
\delta\vev{T^0_{00}}.
\eeq

Before writing the final answer, it is useful to write $\delta\vev{\op}$ in terms of the trace of 
the stress tensor of the deformed theory, 
$T^g$.  The two are related by the dilatation Ward identity, which gives \cite{Osborn1994}
\beq\label{eqn:dTg}
\delta\vev{T^g} = (\D-d)g\delta\vev{\op}.
\eeq
Then, using the definition of the deformed theory's stress tensor (\ref{eqn:Tgabdef}) and 
summing up the contributions (\ref{eqn:dS2Elg}), (\ref{eqn:dS2Tfin}), and (\ref{eqn:dS1gl}),
the total variation of the entanglement entropy at $O(\lambda^1 g^1)$ is 
\beq \label{eqn:dSlg}
\delta S_{\lambda g} = 2\pi \frac{\Omega_{d-2} R^d}{d^2-1}\left[\delta\vev{T^g_{00}}
-\frac1{2\D-d} \delta\vev{T^g}\right] +\delta S^{(2)}_{\mathcal{E},\lambda g} .
\eeq
Since $\delta S^{(2)}_{\mathcal{E},\lambda g}$ is subleading, 
this matches the result (\ref{eqn:Dneqdt}) quoted in the introduction, apart from the $R^{2\D}$
term, which is not present because we have arranged for the renormalized vev 
$\vev{\op}_g$ to vanish.  However, as noted in equation (\ref{eqn:R2Dlambda2}), we
do find such a term at second order in $\lambda$.

\subsection{$\D < \dt$} \label{sec:D<dt}
Extending the above calculation to $\D<\dt$ requires the introduction of one novel
element: a modification of the coupling $g(x)$ to include an IR cutoff.  It is straightforward to 
see why this regulator is needed.  The perturbative calculation of the entanglement entropy
involves integrals of the two point correlator over all of space, schematically of the form
\beq
\int d^dx g(x)\mvev{\Big}{\op(0)\op(x)}_0 = \int d^d x \frac{c_\D g(x)}{x^{2\D}}.
\eeq
If this is cut off at a large distance $L$, the integral scales as $L^{d-2\D}$ (or $\log L$ for 
$\D=\dt$) when the coupling $g(x)$ is constant.  This clearly diverges for $\D\leq\dt$.  

The usual story with IR divergences is that resumming the higher order terms remedies
the divergence, effectively imposing an IR cut off.  Presumably this cut off is set by the scale
of the coupling $L_\text{eff}\sim g^{\frac1{\D-d}}$, but since it arises from higher order 
correlation functions, it may also depend on the details of the underlying CFT.  Although
it may still be possible to compute these IR effects in perturbation theory 
\cite{Zamolodchikov1991, Guida1996, Guida1997}, this goes beyond the techniques employed in the present work.  
However, if we work on length scales small compared to the IR scale, it is possible to
capture the qualitative behavior by simply putting in an IR cut off by hand
 (see \cite{Berenstein2014} for a related approach).  We implement this 
IR cutoff by setting the coupling $g(x)$ to zero when $|\tau|\geq L$.\footnote{This will
work only for $\D>\dt-\frac12$.  For lower operator dimensions, a stronger regulator is needed,
such as a cutoff in the radial direction,
but the only effect this should have is to change the value of $\vev{\op}_g$.}  We may then
express the final answer in terms of the vev $\vev{\op}_g$, which implicitly depends on
the IR regulator $L$.  

The bulk term $\delta S^{(2)}$ involves a new set of divergences from the $\mathcal{T}$ surface
integral that were not present in the original calculation for $\D>\dt$ \cite{Faulkner2015}.  
To compute these divergences and show that they cancel, we will need the real time 
behavior of the bulk scalar fields, in addition to its behavior at $t=0$.  
These are described in appendix \ref{sec:realt}.
The important features are that $\phi_0$ on the $t=0$ surface takes the form
\beq\label{eqn:phi0Etext}
\phi_0 = -\frac{\vev{\op}_g}{2\D-d} z^\D + g z^{d-\D},
\eeq
and the vev $\vev{\op}_g$ is determined in terms of the IR cutoff $L$ by
\beq
\vev{\op}_g = 2gL^{d-2\D}\frac{\Gamma(\D-\dt+\frac12)}{\sqrt{\pi}\,\Gamma(\D-\dt)}.
\eeq
For $t>0$, the time-dependent is given by
\beq\label{eqn:phi0Ttext}
\phi_0 = -\frac{\vev{\op}_g}{2\D-d} z^\D + g z^{d-\D} F(t/z),
\eeq
where the function $F$ is defined in equation (\ref{eqn:F}).  To compute the divergences
along $\mathcal{T}$, the form of this function is needed in the region $t\gg z$, where it simply becomes 
\beq
F(t/z)\xrightarrow{t\gg z} B \left(\frac{t}{z}\right)^{d-2\D},
\eeq
with the proportionality constant $B$ given in equation (\ref{eqn:Fasym}).
The field $\phi_\omega$ behaves similarly as long as $\omega^{-1}\gg z,t$.  In particular,
it has the same form as $\phi_0$ in equations (\ref{eqn:phi0Etext}) and (\ref{eqn:phi0Ttext}),
but with $g$ replaced by $\lambda_\omega$, and $\vev{\op}_g$ replaced with $\delta\vev{\op}$,
given by
\beq
\delta\vev{\op} = \lambda_\omega\frac{2\Gamma(\dt-\D+1)}{\Gamma(\D-\dt)}
\left(\frac{\omega}{2}\right)^{2\D-d},
\eeq
which is the same relation as for $\D>\dt$, equation (\ref{eqn:dvevoD>dt}).

Armed with these solutions, we can proceed to calculate
$\delta S^{(2)}$.  In this calculation, the contribution from the timelike surface $\mathcal{T}$
now has a novel role.  Before, when $\D>\dt$, the integral from this surface died
off as $z\rightarrow 0$ in the region $t_B>z$, and hence the integral there did not need
to be evaluated.  For $\D<\dt$, rather than dying off, this integral is now leads to divergences
as $z\rightarrow0$.  These divergences either cancel among themselves, or cancel against
divergences coming from bulk Euclidean surface $\mathcal{E}$, so that a finite answer is 
obtained in the end.  These new counterterm divergences seem to be related to the 
alternate quantization in holography \cite{Klebanov1999, Casini2016a}, which
invokes a different set of boundary counterterms when defining the bulk AdS action.  It 
would be interesting to explore this relation further.  

At first order in $g$ and $\lambda$, three types of terms will appear,  proportional to each of
$\vev{\op}_g\,\delta\vev{\op}$,  $(g\delta\vev{\op} +\lambda(0)\vev{\op}_g)$, or 
$g \lambda(0)$.  Here, we allow $\lambda(0)\neq0$ because there are no UV divergences 
arising in the energy density or $\op$ expectation values when $\D<\dt$. The descriptions
of the contribution from each of these terms are given below, and the details
of the surface integrals over $\mathcal{E}$ and $\mathcal{T}$  are
contained in appendix \ref{sec:D<dtcalc}.  

The $\vev{\op}_g\delta\vev{\op}$ term has both a finite and a divergent piece coming 
from the integral over $\mathcal{E}$ (see equation (\ref{eqn:dSEz2D})).  This divergence
is canceled by the $\mathcal{T}$ integral in the region $t_B\gg z_0$.  This is interesting since
it differs from the $\D>\dt$ case, where the bulk divergence was canceled by the $\mathcal{T}$
integral in the region $t_B \lesssim z_0$.   The final finite contribution from this term is 
\beq
\delta S^{(2)}_{\mathcal{E},1} = -2\pi \vev{\op}_g\,\delta\vev{\op}
\frac{\Omega_{d-2}}{d^2-1} R^{2\D}
\frac{\D \Gamma(\dt+\frac32)\Gamma(\D-\dt+1)}{ (2\D-d)^2 \Gamma(\D+\frac32)} .
\eeq
It is worth noting that we can perform the exact same calculation with $\vev{\op}_g\delta
\vev{\op}$ replaced by $\frac12\delta\vev{\op}^2$ to compute the second order in $\lambda$
change in entanglement entropy.  The value found in this case agrees with  holographic
results \cite{Casini2016a}.  

The $g\delta\vev{\op}+\lambda(0)\vev{\op}_g$ term receives no contribution from the 
$\mathcal{E}$ surface at leading order since this term in $\phi^2$ scales as $z^d$ 
in the bulk, and the $z$-derivatives in the Laplacian $\nabla^2$ annihilate such a term.  
The surface $\mathcal{T}$ produces a finite term, plus a collection of divergent terms 
from both regions $t\sim z$ and $t\gg z$, which cancel among themselves.  The finite 
term is given by
\beq
\delta S^{(2)}_{\mathcal{T},2} = 2\pi \frac{\Omega_{d-2} R^d\D}{(d^2-1)(2\D-d)}(g\delta\vev{\op}
+\lambda(0)\vev{\op}_g), 
\eeq
which is exactly analogous to the term (\ref{eqn:dS2Tfin}) found for the case $\D>\dt$.  

Finally, the term with coefficient $\lambda(0) g$ produces subleading terms, scaling as 
$R^{2(d-\D+n)}$ for positive integers $n$.  Since these terms are subleading, we do not
focus on them further.  In this case, it must also be shown that the divergences appearing
in the $\mathcal{T}$ cancel amongst themselves, since no divergences arise from the 
$\mathcal{E}$ integral.  The calculations in appendix \ref{sec:D<dtcalc} verify that 
this indeed occurs.

We are now able to write down the final answer for the change in entanglement entropy for
$\D<\dt$.  The contribution from $\delta S^{(1)}$ is exactly the same as the $\D>\dt$ case,
and is given by (\ref{eqn:dS1gl}).  Following the same steps that led to equation 
(\ref{eqn:dSlg}), the contributions from the finite piece of $\delta S^{(2)}_{\mathcal{E},1}$ 
in (\ref{eqn:dSEz2D}) and $\delta S^{(2)}_{\mathcal{T},2}$ in (\ref{eqn:dS2T2}) combine
with $\delta S^{(1)}$ to give
\beq
\delta S_{\lambda g} = \frac{2\pi \Omega_{d-2}}{d^2-1} \left[R^d\left(
\vev{T^g_{00}} -\frac{1}{2\D-d}
\vev{T^g} \right)-R^{2\D} \vev{\op}_g\delta\vev{\op} \frac{\D\Gamma(\dt+\frac32) 
\Gamma(\D-\dt+1)}{(2\D-d)^2 \Gamma(\D+\frac32)}   \right],
\eeq
where we have set $\lambda(0)=0$ for simplicity and to match the expression for 
$\D>\dt$, which required $\lambda(0)=0$.  

\subsection{$\D = \dt$} \label{sec:D=dt}
Similar to the $\D<\dt$ case, there are IR divergences that arise when $\D=\dt$.  These 
are handled as before with an IR cutoff $L$, on which the final answer explicitly 
depends.  A new feature arises, however, when expressing the answer in terms of  $\vev{\op}_g$
rather than $L$: the appearance of a renormalization scale $\mu$.  The need 
for this renormalization scale can be seen by examining the expression for $\vev{\op}_g$, 
which depends on the $\op\op$ two-point function with $\D=\dt$:
\beq
\vev{\op}_g = -\int d^d x \frac{g c'_\D}{x^d}  =
-gc'_\D\frac{\pi^{\dt}}{\Gamma(\dt)} \int \frac{d\tau}{\tau}. 
\eeq
This has a logarithmic divergence near $x=0$ which must be regulated.  The UV-divergent
piece can be extracted using the point-splitting cutoff for $|\tau|<\delta$; however, there is
an ambiguity in identifying this divergence since the upper bound of this integral
cannot be sent to $\infty$.  The appearance of the renormalization scale is related to 
matter conformal anomalies that exist for special values of $\D$ \cite{Osborn:1991gm,
Osborn1994, Petkou1999}.   Thus we must impose an upper cutoff on the integral, which
introduces the renormalization scale $\mu^{-1}$.  The divergent piece of $\vev{\op}_g$
is then
\beq
\vev{\op}_g^\text{div.} = g c'_\D \frac{\pi^{\dt}}{\Gamma(\dt)} 2\log\mu\delta.
\eeq
Now we can determine the renomalized vev of $\op$, using the IR-regulated $\tau$
integral,
\begin{align}
\vev{\op}_g^\text{ren.} = \vev{\op}_g-\vev{\op}_g^\text{div.} &= - \int^L d\tau\int d^{d-1}x
\frac{g c'_\D}{x^d} - gc'_\D \frac{\pi^{\dt}}{\Gamma(\dt)} 2\log\mu\delta \\
&=
\label{eqn:vevOren}
-gc'_\D\frac{\pi^{\dt}}{\Gamma(\dt)} 2\log \mu L.
\end{align}
The final answer we derive for the entanglement entropy when $\D=\dt$ will depend on 
$\log L$ but not on explicitly $\mu$ or $\vev{\op}_g$.  Only after rewriting it in terms of 
$\vev{\op}^\text{ren.}_g$ does the $\mu$ dependence appear.  

One other small modification is necessary when $\D=\dt$.  The normalization  $c_\D$
for the $\op\op$ two point function defined in (\ref{eqn:cD}) has a double zero at $\D=\dt$ 
which must be removed.  
This is easily remedied by dividing by $(2\D-d)^2$ \cite{Klebanov1999, Freedman:1998tz}, 
so that 
the new constant appearing in the two point function is 
\beq \label{eqn:cD'}
c'_\D = \frac{\Gamma(\D)}{2\pi^{\dt} \Gamma(\D-\dt+1)} \;\xrightarrow{\hphantom{n}
\D\rightarrow\dt \hphantom{n}} \;
\frac{\Gamma(\dt)}{2\pi^{\dt}} .
\eeq
This change affects the normalization of the bulk field $\phi$ by dividing by a single
factor of $1/(2\D-d)$, so that
\beq\label{eqn:phicD'}
\phi(x_B) = \frac{\Gamma(\dt)}{2\pi^{\dt}}\int_{C(\delta)} d\tau
\int d^{d-1}\vec{x}
\frac{z^\D f(\tau,\vec{x})}{(z^2+(\tau - i t_B)^2+ (\vec{x}-\vec{x}_B)^2)^\D}.
\eeq

These are all the components needed to proceed with the calculation of the entanglement 
entropy.  As before, we solve for the bulk field $\phi_0$ associated with a constant 
coupling $g$, set to zero for $|\tau|>L$.  The $\phi_\omega$ field associated with the 
state deformation $\lambda = \lambda_\omega \cos\omega\tau$ is again given by 
a modified Bessel function on the Euclidean section.  Its form along the timelike surface
$\mathcal{T}$ is derived from the integral representation (\ref{eqn:phicD'}), and particular
care must be taken in the region $t_B\sim z$, where a divergence in $\delta$ appears.
Although this divergence is not present if we require $\lambda(0)=0$, we analyze the 
terms that it produces for generality.  This $\delta$ divergence is shown to cancel
against a similar divergence in $\delta S^{(1)}$ related to the divergence in the $\vev{T_{00}
\op\op}$ three-point function.   

The full real-time solutions for $\phi_0$ and $\phi_\omega$ are given 
in appendix \ref{sec:D=dtrt}.
The $\phi_0$ solution from equation (\ref{eqn:gzd2G}) takes the form
\beq \label{eqn:phi0gG}
\phi_0 = gz^{\dt} G(t_B/z, \delta / z, L/z),
\eeq 
with the function $G$ defined in equation (\ref{eqn:G}).  The dependence of this 
function on $\delta$ is needed only in the region $t_B\sim z$; everywhere else it can
safely be taken to zero.  On the $\mathcal{E}$ surface where 
$t_B=0$, the solution  in the limit $L\gg z$ is 
\beq \label{eqn:phi0vevOren}
\phi_0 = g z^{\dt} \log\frac{2L}{z}  = -\vev{\op}_g^\text{ren.} - g z^{\dt} \log \frac{\mu z}{2},
\eeq 
where the second equality uses the value of $\vev{\op}_g^{\text{ren.}}$ derived in 
(\ref{eqn:vevOren}).  
We also need $\phi_0$ in the region $t_B\gg z$, given by
\beq
\phi_0 = g z^{\dt}\log \frac{L}{t_B} .
\eeq
For $\phi_\omega$, the solution on the $\mathcal{E}$ surface is still given by a modified Bessel
function as in equation (\ref{eqn:phio}), but must be divided by $(2\D-d)$ according 
to our new normalization, 
\beq
\phi_\omega = \lambda_{\omega} z^\dt K_0(\omega z)\xrightarrow{z\rightarrow 0}-
\lambda_\omega z^{\dt} \left(
\gamma_E + \log\frac{\omega z}{2}\right). 
\eeq  
By writing the argument of the $\log$ term as in equation (\ref{eqn:phi0vevOren}), one can
read off the renormalized operator expectation value,
\beq \label{eqn:dOrentext}
\delta\vev{\op}^\text{ren.} = \lambda_\omega \left(\gamma_E + \log\frac{\omega}{\mu}\right).
\eeq
Beyond $t_B=0$, as long as $\omega^{-1}\gg t_B$, the solution can be written in a similar
form as (\ref{eqn:phi0gG}).  This is given by equation (\ref{eqn:phiologot}), which 
reduces when $t_B\gg z$ to
\beq
\phi_\omega = -\lambda_\omega z^{\dt}(\gamma_E + \log \omega t_B). 
\eeq

Now that we have the form of the solutions on the surfaces $\mathcal{E}$ and $\mathcal{T}$, 
the entanglement calculation contains four parts.  The first is the integral over $\mathcal{E}$,
where a $\log z_0$ divergence appears.  This cancels against a collection of divergences
from the $\mathcal{T}$ surface.  The second part is the $\mathcal{T}$ surface near 
$t_B\sim z$.  This region produces  more divergences
in $z_0$ and $\delta$, some of which cancel the bulk divergence.  
The third part is the integral over $\mathcal{T}$ for $t_B\gg z$, 
which eliminates the remaining $z_0$ divergences.  Finally, an additional divergence from
the stress tensor in $\delta S^{(1)}$ cancels the  $\delta$ divergence, producing a finite
answer.

Appendix \ref{sec:D=d2calc} describes the details of these calculations.  In the end, 
the contributions from 
equations (\ref{eqn:dS2Efinlog}), (\ref{eqn:dS2Ediv}),
(\ref{eqn:dS2Tdivlog}), (\ref{eqn:dS2T1+2}) and (\ref{eqn:deltact}) combine
together to give the following total change in entanglement entropy, at $O(\lambda^1 g^1)$,
\begin{align}
\delta S_{\lambda_\omega g}&=2\pi  \frac{\Omega_{d-2}R^d}{d^2-1}\left\{
\delta\vev{T^0_{00}}^\text{ren.} +g\lambda_\omega \left[ 
\dt \logp{\frac{2L}{R}}
\left(\gamma_E + \log\frac{\omega R}{2}\right)    \right. \right. \nonumber \\
&\left.\left. + \frac{d}{4}H_{\frac{d+1}{2}}
\left(\gamma_E+\log\frac{R^2\omega}{4L}\right)
 -  \log\mu R  -\frac18
\left(H^{(2)}_{\frac{d+1}{2}} +H_{\frac{d+1}{2}} (H_{\frac{d+1}{2}}-2)\right) \right]
\right\}.    \label{eqn:D=d2final}
\end{align}
This is the answer for a single frequency $\omega$ in the state deformation function 
$\lambda(x)$.  Since $\lambda(0)\neq 0$, this result cannot be immediately interpreted as 
the entanglement entropy of an excited state, since the state has a divergent expectation 
value for $\op$.\footnote{However, viewing $\omega$ as an IR regulator, this equation can be 
adapted to express
the change in \emph{vacuum} 
entanglement entropy between a CFT and the deformed theory.}
To get the entanglement entropy for an excited state, we should integrate over all 
frequencies, and use the fact that $\int d\omega \lambda_\omega=0$.  When this is 
done, all terms with no $\log\omega$ dependence drop out.  Also, we no longer need to specify
that operator expectation values are renormalized, since the change in expectation values
between two states is finite and scheme-independent.  

We would like to express the answer in terms of $\delta\vev{\op}$.  By integrating equation 
(\ref{eqn:dOrentext}) over all frequencies and using that $\lambda(0)=0$, we find
\beq
\delta\vev{\op} = \int_0^\infty d\omega\,\lambda_\omega\log\omega.
\eeq
With this, the total change in entanglement entropy for nonsingular states coming from 
integrating \ref{eqn:D=d2final} over all frequencies is 
\beq
\delta S_{\lambda g} = 2\pi\frac{\Omega_{d-2} R^d}{d^2-1} \left[
\delta\vev{T^0_{00}} + g\dt\delta\vev{\op} \left(\frac12 H_{\frac{d+1}{2}}+\log\frac{2L}{R}\right)
\right].
\eeq
This can be expressed in terms of the deformed theory's stress tensor $T^g_{00}$ and 
trace $T^g$ using equations (\ref{eqn:Tgabdef}) and (\ref{eqn:dTg}),
\beq \label{eqn:D=d2logL}
\delta S_{\lambda g} = 2\pi\frac{\Omega_{d-2} R^d}{d^2-1} \left[
\delta\vev{T^g_{00}} + \delta\vev{T^g}\left(\frac2d-\frac12 H_{\frac{d+1}{2}} + \log\frac{R}{2L}
\right)
\right].
\eeq
Although the answer is 
scheme-independent in the sense that $\mu$ does not explicitly appear, there is a dependence
on the IR cutoff $L$.  This cutoff is related to the renormalized vev $\vev{\op}^\text{ren.}_g$
via (\ref{eqn:vevOren}), which does depend on the renormalization scheme.  
Thus the dependence on $L$ in
the above answer can be traded for $\vev{\op}_g^\text{ren.}$, at the cost of introducing 
(spurious) $\mu$-dependence, 
\beq \label{eqn:dSlgvevO}
\delta S_{\lambda g} = 2\pi\frac{\Omega_{d-2} R^d}{d^2-1} \left[
\delta\vev{T^g_{00}} + \delta\vev{T^g}\left(\frac2d-\frac12 H_{\frac{d+1}{2}} + \log\frac{\mu R}{2}
\right) -\dt \vev{\op}_g\delta\vev{\op} 
\right],
\eeq
which is the result quoted in the introduction, equation (\ref{eqn:Deqdt}).

\section{Discussion} \label{sec:discussion}

The equivalence between the Einstein equation and maximum vacuum entanglement 
of small balls relies on 
a conjecture about the behavior of the entanglement entropy of excited states, equation
(\ref{eqn:dSIRmod}).  This work has sought to check the conjecture in CFTs deformed
by a relevant operator.  In doing so, we have derived new results on the behavior of
excited state entanglement entropy in such theories, encapsulated by equations
(\ref{eqn:Dneqdt}) and (\ref{eqn:Deqdt}).  These results agree with holographic calculations
\cite{Casini2016a}
that employ the Ryu-Takayanagi formula.  Thus, this work extends those results to 
\emph{any} CFT, including those which are not thought to have holographic duals.  

For deforming operators of dimension $\D>\dt$ considered in section \ref{sec:D>dt}, 
the calculation is a straightforward 
application of Faulkner's method for computing entanglement entropies \cite{Faulkner2015}.  
One subtlety in this case is the presence of UV divergences in 
$\delta\vev{\op}$ and $\delta\vev{T^0_{00}}$ unless
the state deformation function $\lambda(x)$ is chosen appropriately.  As discussed in 
section \ref{sec:excited}, this translates to the condition that $\lambda$ and sufficiently
many of its $\tau$-derivatives vanish on the $\tau=0$ surface.  When the entanglement
entropy of the state is calculated, this condition implies that terms scaling with the ball
radius as $R^{2(d-\D+n)}$, which are present for generic $\lambda(x)$, vanish, 
where $n$ is a positive integer less than or equal to
$\left\lfloor\D-\dt\right\rfloor$.  As $R$ approaches
zero, these terms  
dominate over the energy density term, which scales as $R^d$.  This shows that regularity 
of
the state translates to the dominance of the modular Hamiltonian term 
in the small ball limit when $\D>\dt$.  The subleading terms arising from 
this calculation are  given in equation (\ref{eqn:dS2Elg}).

Section \ref{sec:D<dt} then extends this result to operators of dimension $\D<\dt$.  In this 
case, IR divergences present a novel facet to the calculation.  To deal with these divergences,
we impose an IR cutoff on the coupling $g(x)$ at  scale $L$.  A more complete 
treatment of the IR divergences would presumably involve resumming higher order 
contributions, which then would effectively impose an IR cutoff in the lower order terms.  
This cutoff should be of the order $L_\text{eff.}\sim g^{\frac{1}{\D-d}}$, but can depend
on other details of the CFT, including any large parameters that might be present.  Note
this nonanalytic dependence of the IR cutoff on the coupling signals nonperturbative
effects are at play \cite{Nishioka:2014kpa,Herzog2013}.  
After the IR cutoff is imposed, the calculation
of the entanglement entropy proceeds as before.  In the final answer, equation 
(\ref{eqn:Dneqdt}), the explicit dependence on the IR cutoff is traded for the renormalized
vacuum expectation value $\vev{\op}_g$.  This expression agrees with the holographic
calculation to first order in $\delta\vev{\op}$ in the case that $\vev{\op}_g$ is nonzero
\cite{Casini2016a}.  

Finally, the special case of $\D=\dt$ is addressed in section \ref{sec:D=dt}.  Here, both
 UV and IR divergences arise, and these are dealt with in the same manner as the 
$\D>\dt$ and $\D<\dt$ cases.  The answer before imposing that the state is nonsingular
is given in equation (\ref{eqn:D=d2final}), and it depends logarithmically on an arbitrary
renormalization scale $\mu$.   This scale $\mu$ arises when renormalizing the 
stress tensor expectation value $\delta\vev{T^0_{00}}$,  
as is typical of logarithmic UV divergences.  
Note that the dependence on $\mu$ in the final answer is only superficial, since the 
combination $\delta\vev{T^0_{00}}^\text{ren.} - \log\mu{R}$ appearing there is
independent of the choice of $\mu$.  Furthermore, for regular states, $\delta\vev{T^0_{00}}$
is UV finite, and hence the answer may be written without reference to the renormalization scale as in (\ref{eqn:D=d2logL}),
although it explicitly depends on the IR cutoff.  In some cases, such as free field theories, 
the appropriate IR cutoff may be calculated exactly \cite{Casini2009,Blanco2011, 
Casini2016a}.  
Re-expressing the answer in terms of  $\vev{\op}_g$ instead of the IR cutoff, 
as in equation (\ref{eqn:Deqdt}),
re-introduces the renormalization scale $\mu$, since the 
vev requires renormalization and hence is $\mu$-dependent.  Again, this dependence
on $\mu$ is superficial; it cancels between $\vev{\op}_g$ and the $\log\frac{\mu R}{2}$ 
terms.

\subsection{Implications for the Einstein equation} \label{sec:EEimps}
We now ask whether the results (\ref{eqn:Dneqdt}) and (\ref{eqn:Deqdt}) are consistent
with the conjectured form of the entanglement entropy variation (\ref{eqn:dSIRmod}).  The
answer appears to be yes, with the following caveat: the scalar function $C$ explicitly
depends on the ball size $R$.  This comes about from the $R^{2\D}$ in equation 
(\ref{eqn:Dneqdt}), in which case $C$ contains
a piece scaling as $R^{2\D-d}$,  and from the $R^d \log R$ term in (\ref{eqn:Deqdt}),
which gives $C$ a $\log R$ term.  When $\D\leq\dt$, these terms are the dominant
component of the entanglement entropy variation when the ball size is taken to be small.  

The question now shifts to whether $R$-dependence in the function $C$ still allows the 
derivation of the Einstein equation to go through.  As long as $C(R)$ transforms as a 
scalar under Lorentz boosts for fixed ball size $R$, the tensor equation (\ref{eqn:EEtensor})
still follows from the conjectured form of the entanglement entropy variation (\ref{eqn:dSIRmod})
\cite{Jacobson2015}.  
One then concludes from stress tensor conservation and the Bianchi identity that the
curvature scale of the maximally symmetric space characterizing the local vacuum
is dependent on the size of the ball, $\Lambda=\Lambda(x,R)$.\footnote{This idea 
was proposed by Ted Jacobson, and I thank him for 
for discussions regarding this point.}  
There does not seem to be an 
immediate reason  disallowing an $R$-dependent $\Lambda$.  

There are two requirements on $\Lambda(R)$ for this to be a valid interpretation.  First, 
$\Lambda^{-1}$ should remain much larger than $R^2$ in order to justify using the 
flat space conformal Killing vector (\ref{eqn:CKV}) for the CFT modular Hamiltonian, and 
also to justify keeping only the first order correction to the area due to 
curvature in equation (\ref{eqn:dSUVG}).  
Since $C(R)$ is dominated by the $R^{2\D}$ for $\D\leq\dt$ as $R
\rightarrow 0$, it determines $\Lambda(R)$ by
\beq
\Lambda(R) = \frac{2\pi}{\eta} C\sim \ell_P^{d-2}\vev{\op}_g\delta\vev{\op} R^{2\D-d}.
\eeq
The the requirement that $\Lambda(R) R^2\ll1$ becomes
\beq\label{eqn:Rupper}
 \frac{R}{\ell_P} \ll \left(\frac{1}{\ell_P^{2\D}\vev{\op}_g\delta\vev{\op}}\right)^{\frac{1}{2\D-d+2}}.
\eeq
Since $2\D-d+2\geq0$ by the CFT unitarity bound for scalar operators, this 
inequality can always be satisfied
by choosing $R$ small enough.  Furthermore, since $\vev{\op}_g\delta\vev{\op}$ should 
be  small in Planck units, the right hand side of this inequality is large, and hence can be 
satisfied for $R\gg\ell_P$.   A second requirement is that $\Lambda$ remain sub-Planckian
to justify using a semi-classical vacuum state when discussing the variations.  
This means $\Lambda(R)
\ell_P^2 \ll1$, which then implies
\beq\label{eqn:Rlower}
\frac{R}{\ell_P} \gg \Big(\ell_P^{2\D}\vev{\op}_g\delta\vev{\op}\Big)^{\frac{1}{d-2\D}}
\eeq
This now places a lower bound on the size of the ball for which the derivation is valid.  However,
the $R$-dependence in $\Lambda(R)$ is only significant when $d-2\D$ is positive, and hence
the right hand side of this inequality is small.  Thus, there should be a wide range of $R$ values
where both (\ref{eqn:Rupper}) and (\ref{eqn:Rlower}) are satisfied.  
The implications of such an $R$-dependent 
local curvature scale merits further investigation.  Perhaps it is related to 
a renormalization group flow of the cosmological constant \cite{Shapiro2009}.

A second, more speculative possibility is 
that the $R^{2\D}$ and $\log R$ terms are resummed due to higher
order corrections into something that is subdominant in the $R\rightarrow 0$ limit.  One reason
for suspecting that this may occur is that the $R^{2\D}$ at second order in the state variation
can dominate over the lower order $R^d$ terms at small $R$, possibly hinting at a break
down of perturbation theory.\footnote{However, reference \cite{Casini2016a} found that terms
at third order in the state variation are subdominant to this term for small values of $R$.} 
As a trivial example, 
suppose the $R^{2\D}$ term arose from a function of the form
\beq
\frac{R^d}{1+(R/R_0)^{2\D-d}}.
\eeq
Since $\D<\dt$, this behaves like $R^d-R^{2\D}R_0^{d-2\D}$ 
when $R\gg R_0$.  
However, about $R=0$, 
it becomes
\beq
\frac{R^d}{1+(R/R_0)^{2\D-d}}\xrightarrow{R\rightarrow 0} R_0^d\left(\frac{R}{R_0}\right)^{2
(d-\D)},
\eeq
which is subleading with respect to a term scaling as $R^d$.  Note however that something must 
determine the scale $R_0$ in this argument, and it is difficult to find a scale that is free of
nonanalyticities in the coupling or operator expectation values.  It would be interesting to
analyze whether these sorts of nonperturbative effects  play a role in the entanglement
entropy calculation.  

Finally, one may view the $R$ dependence in $\Lambda$ as evidence that the relation
between maximal vacuum entanglement and the Einstein equation does not hold for some 
states.  In fact, there is some evidence that the relationship must not hold for some states for
which the entanglement entropy is not related to the energy density of the state.  A 
particular example is a coherent state, which has no additional entanglement entropy 
relative to the vacuum despite possessing energy \cite{Varadarajan2016}.

\subsection{Future work}
This work leads to several possibilities for future investigations.  
First is the question
of how the entanglement entropy changes under a change of Lorentz frame.  The 
equivalence between vacuum equilibrium and the Einstein equation rests crucially on 
the transformation properties of the quantity $C$ appearing in equation (\ref{eqn:dSIRmod}).  
Only if it transforms as a scalar can it be absorbed in to the local curvature scale $\Lambda(x)$.  
The calculation in this work was done for a large class of states 
defined by Euclidean path integral.
For a boosted state, one could 
simply repeat the calculation using the Euclidean space relative to the boosted frame, and 
the same form of the answer would result.  For states considered here that were stationary
on time scales on the order $R$ (since $\omega R\ll1$), it seems plausible that the states 
constructed in the boosted Euclidean space contain the boosts of the original states. However,
this point should be investigated more thoroughly.  
Another possibility for checking how the entanglement entropy changes under boosts is to use
the techniques of \cite{Faulkner2015a}, which perturbatively evaluates the 
change in entanglement entropy under a deformation of the region $\Sigma$.  In particular, 
they derive a formula that applies for timelike deformations of the surface, and hence 
could be used to investigate the behavior under boosts.  

Performing the calculation to the next order in perturbation theory would also provide 
new nontrivial checks on the conjecture, in addition to providing new insights for the 
general theory of perturbative entanglement entropy calculations.  This has been 
done in holography \cite{Casini2016a}, so it would be interesting to see if the holographic
results continue to match for a general CFT.  The entanglement entropy at the next order in 
perturbation theory depends on the $\op\op\op$ three point function \cite{Rosenhaus:2014ula}. 
One reason for suspecting that the holographic results still match
stems from the universal form of this three point function in CFTs. 
For
scalar operators, it is completely fixed by conformal invariance up to an overall constant.  
Thus, up to the multiplicative constant in the three-point function, 
there is nothing in the calculation distinguishing 
between holographic and non-holographic theories.  
At higher order, one would eventually expect the 
holographic calculation to differ from the general case.  For example, the four point function
has much more freedom, depending on an arbitrary function of two conformally invariant
cross-ratios.  It is likely that universal statements about the entanglement entropy would 
be hard to make at that order.

The IR divergences when $\D\leq\dt$ were dealt with using an IR cutoff, which  
captures the qualitative behavior of the answer, but misses out on the precise details of 
how the coupling suppresses the IR region.  It may be possible to improve on this calculation
at scales above the IR scale using established techniques for handling IR divergences 
perturbatively \cite{Zamolodchikov1991,Guida1996,Guida1997}, or by examining
specific cases that are exactly solvable \cite{Casini2009,Blanco2011,Zamolodchikov1991}.  
IR divergences continue to plague the calculations at 
 higher order in perturbation theory.  This   
can be seen by examining the $\op\op\op$ three point function,
\begin{align}
\iint d^d x_1 d^d x_2 \mvev{\big}{\op(0)\op(x_1)\op(x_2)} = \iint d^d x_1 d^d x_2
 \frac{c}{|x_1|^\D |x_2|^\D|x_1-x_2|^\D}. 
\end{align}
By writing this in spherical coordinates, performing the angular integrals, and 
defining $u = \frac{r_2}{r_1}$, this may be written
\beq
c\Omega_{d-1} \Omega_{d-2} \pi \int_0^\infty du \int_0^\infty dr_1\, r_1^{2d-3\D-1} u^{d-\D-1}
(1+u)^{-\D} \tensor[_2]{F}{_1}\left(\frac12,\frac\D2;1;\frac{2u}{(1+u)^2}\right),
\eeq
This is clearly seen to diverge in the IR region $r_1\rightarrow\infty$ when $\D\leq\frac{2d}{3}$,
so that some operators that produced IR finite results in the two-point function now 
produce IR divergences.

Finally, one may be interested in extending Jacobson's derivation to include higher
order corrections to the Einstein equation.  On the geometrical side, this involves considering
higher order terms in the Riemann normal coordinate expansion of the metric about a point.  
This could also lead to deformations of the entangling surface $\partial \Sigma$, and these effects
could be computed perturbatively using the techniques of \cite{Rosenhaus2014,
Rosenhaus:2014ula, Rosenhaus:2014zza, Faulkner2015a}.  It may be interesting to 
see whether these expansions can be carried out further to compute the higher curvature
corrections to Einstein's equation.

\acknowledgments
It is a pleasure to thank Anton de la Fuente, Tom Faulkner, Dami\'{a}n Galante, 
Sungwoo Hong, Ted Jacobson, Rob Myers, Vladimir Rosenhaus, and Raman Sundrum
for helpful discussions.  I am especially grateful to Horacio Casini, Dami\'{a}n Galante
and Rob Myers for correspondence on their related work and for sharing early results with me,
and to Ted Jacobson for comments on a draft of this work.  
I thank the organizers of the programs ``Quantum Gravity Foundations: UV to IR" and 
``Entanglement for Strongly Correlated Matter" held at the Kavli Institute for Theoretical
physics, and thank the KITP and the Perimeter Institute for Theoretical Physics for hospitality
during work on this project.  This research was supported in part by the 
National Science Foundation under grants No.\ PHY-1407744 and PHY11-25915, and
by Perimeter Institute for Theoretical Physics.  Research at Perimeter Institute is supported 
by the Government of Canada through Industry Canada and by the Province of Ontario through 
the Ministry of Research and Innovation.

\appendix

\section{Coefficients for the bulk expansion} \label{sec:Bessel}
This appendix lists the coefficients appearing in section \ref{sec:D>dt} for the expansion of
$\phi_\omega$ and $\nabla^2 \phi_0\phi_\omega$.  Given its definition (\ref{eqn:phio}), the
coefficients appearing in the expansion (\ref{eqn:phioseries}) follow straightforwardly from
known expansions of the modified Bessel functions \cite{NIST:DLMF}:
\begin{align}
a_n &= \frac{\Gamma(\dt-\D+1)}{4^n\, n!\Gamma(\dt-\D+n+1)} \\
b_n &=\frac{\Gamma(\D-\dt+1)}{4^n\, n!\Gamma(\D-\dt+n+1)}.
\end{align}
When acting with $\nabla^2$ on the series $\phi_0\phi_\omega$, the $\tau$ and $z$ derivatives
mix adjacent terms in the series.  The relation this gives is 
\beq
c_n = 2(d-\D+n)(d-2\D+2n)a_n-a_{n-1},
\eeq
which, given the properties of the $a_n$, simplifies to
\beq
c_n = 2(d-\D)(d-2\D+2n)a_n.
\eeq
Similarly, for the $d_n$ series,
\beq
d_n = 2n(d+2n) b_n - b_{n-1},
\eeq
which implies
\beq
d_n = 4n(d-\D) b_n.
\eeq

\section{Real-time solutions for $\phi(x)$}

\subsection{$\D<\dt$} \label{sec:realt}

This appendix derives the real time behavior of the fields $\phi_0$ and $\phi_\omega$.  Starting
with $\phi_0$, the coupling $g(x)$ is a constant $g$ for $|\tau|$ less than
the IR cutoff $L$, 
and zero otherwise. The bulk solution  found by evaluating (\ref{eqn:phiint}) is
\begin{align}
\label{eqn:phi0rtint}
\phi_0 &= g z^{d-\D}\frac{\Gamma(\D-\dt+\frac12)}{\sqrt{\pi}\,\Gamma(\D-\dt)}\left[\int_0^{L/z} dy\,
\left(1+(y-it_B/z)^2\right)^{\dt-\D-\frac12} + \text{c.c.}\right]  \\
&=  g z^{d-\D}\frac{\Gamma(\D-\dt+\frac12)}{\sqrt{\pi}\,\Gamma(\D-\dt)}\left[ 
\frac{L-i t_B}{z} \; \tensor[_2]{F}{_1}\left(\frac12, \D-\dt+\frac12; \frac32;
\frac{-(L-i t_B)^2}{z^2} 
\right)   \right. \nonumber \\
&\qquad\qquad \left.  +\frac{ it_B}{z}\; \tensor[_2]{F}{_1}\left(\frac12, \D-\dt +\frac12; \frac32;
\frac{t_B^2}{z^2} \right)  +\text{c.c.}\right].  \label{eqn:Lisigma}
\end{align}
Here, notice that no cut off near  $y=0$
was needed, since the $\op\op$ two point function
has no UV divergences.  However, one still has to be mindful of the branch prescription,
which is appropriately handled by adding the complex conjugate as directed in the expressions
above (denoted by ``c.c.'').  When $t_B>z$, the branch in the hypergeometric function 
along the real axis is dealt with by replacing $t_B\rightarrow t_B+i \delta$, and taking the 
$\delta\rightarrow 0$ limit.  

This solution can be simplified in the two regimes of interest, namely on 
$\mathcal{E}$ with $t_B=0$ and on $\mathcal{T}$ in the $z\rightarrow 0$ limit.  In the first 
case, $\phi_0$ reduces to
\beq\label{eqn:phi0euclidean}
\phi_0\big|_{t_B=0} = g z^{d-\D} - z^\D\, \frac{g L^{d-2\D} \Gamma(\D-\dt+\frac12)}{
\sqrt{\pi}\, \Gamma(\D-\dt+1)} 
\; \tensor[_2]{F}{_1}\left(\D-\dt, \D-\dt+\frac12; \D-\dt +1;
\frac{-z^2}{L^2}\right),
\eeq
and since we are assuming $R\ll L$, we only need this in the small $z$ limit, 
\beq\label{eqn:phi0slz}
\phi_0\rightarrow g z^{d-\D} - z^\D\, \frac{g L^{d-2\D} \Gamma(\D-\dt+\frac12)}{
\sqrt{\pi}\, \Gamma(\D-\dt+1)} .
\eeq
From this, one immediately reads off the vev of $\op$,
\beq
\vev{\op}_g = 2 gL^{d-2\D}\frac{\Gamma(\D-\dt+\frac12)}{\sqrt{\pi}\,\Gamma(\D-\dt)}.
\eeq
The real time behavior near $z\rightarrow0$ and with $t_B\ll L$ takes the form 
\beq \label{eqn:phi0realt}
\phi_0 = -\frac{\vev{\op}_g}{2\D-d}z^\D + g z^{d-\D} F(t_B/z),
\eeq
with 
\beq \label{eqn:F}
F(s) = \begin{cases}
\hfil 1&s<1 \\
 \frac{\sqrt{\pi}\, (s^2-1)^{\dt-\D +\frac12}}{s\, \Gamma(\D-\dt+1)\, \Gamma(\dt-\D+\frac12)}\;\;
\tensor[_2]{F}{_1}\left(1,\frac12;\D-\dt+1;\frac{1}{s^2}\right) & s>1
\end{cases} .
\eeq
In particular, for large argument, this function behaves as
\beq \label{eqn:Fasym}
F(s\rightarrow\infty) =Bs^{d-2\D};\qquad 
B= \frac{\sqrt{\pi}}{\Gamma(\D-\dt+1)\Gamma(\dt-\D+\frac12)}.
\eeq

We also need the solution for the field corresponding to the state deformation $\lambda(x)$.  
The oscillatory behavior for the choice (\ref{eqn:lambda}) for this function serves to
regulate the IR divergences, and hence no additional IR cutoff is needed.  Thus the 
bulk solution on the Euclidean section (\ref{eqn:phio}) is still valid.  The real time
behavior of the solution is given by the following integral,
\beq
\phi_\omega = \lambda_\omega z^{d-\D}\frac{\Gamma(\D-\dt+\frac12)}{\sqrt{\pi}\,
\Gamma(\D-\dt)}\left[\int_0^{\infty} dy\,\cos(\omega z y)
\left(1+(y-it_B/z)^2\right)^{\dt-\D-\frac12} + \text{c.c.}\right] .
\eeq
To make further progress on this integral, we note that we only need the solution
up to times $t_B\sim R\ll\omega^{-1}$.  In this limit, the solution 
should not be sensitive to the details
of the IR regulator. Thus, the  answer should be the same as for $\phi_0$ in (\ref{eqn:phi0realt}),
the 
only difference being the numerical value for the operator expectation value.  
This behavior can be seen by breaking the integral into two regions, 
$(0,\frac{a}{z})$ and $(\frac{a}{z},\infty)$, with
$t_B\ll a \ll \omega^{-1}$.  In the first region, the cosine can be set to $1$ since
its argument is small.  The resulting integral is identical to (\ref{eqn:phi0rtint}), with $L$
replaced by $a$.  
In the second region, the integration variable $y$ is large compared to $1$ and $t_B/z$,
so the integral reduces to
\begin{align}
\label{eqn:asplit}
&\lambda_\omega z^{d-\D}\frac{2\Gamma(\D-\dt+\frac12)}{\sqrt{\pi}\,
\Gamma(\D-\dt)} \int_{a/z}^{\infty} dy\,\cos(\omega z y) y^{d-2\D-1}  \\
\label{eqn:oa}
=&\; \lambda_\omega z^\D \left(\frac{\omega}{2}\right)^{2\D-d}\, 
\frac{\Gamma(\dt-\D)}{\Gamma(\D-\dt)}  + \lambda_\omega z^{d-\D} \left(\frac{a}{z}\right)^{d-2\D}
\frac{\Gamma(\D-\dt+\frac12)}{\sqrt{\pi}\,
\Gamma(\D-\dt+1)},
\end{align}
valid for $a\ll\omega^{-1}$.  The second term in this expression cancels against the same
term appearing in the first integration region, effectively replacing it with the first term in
(\ref{eqn:oa}).  The final answer for the real time behavior of $\phi_\omega$ near $z=0$
is
\beq \label{eqn:phiort}
\phi_\omega =-\frac{\delta\vev{\op}}{2\D-d}z^\D
+\lambda_\omega z^{d-\D} F(t_B/z).
\eeq
where we have identified $\delta\vev{\op}$ as 
\beq
\delta\vev{\op} = 
\lambda_\omega
\frac{2\,\Gamma(\dt-\D+1)}{\Gamma(\D-\dt)} \left(\frac{\omega}{2}\right)^{2\D-d}.
\eeq

\subsection{$\D=\dt$}  \label{sec:D=dtrt}
Here we derive the real-time behavior of $\phi_0$ and $\phi_\omega$ when $\D=\dt$. 
We begin with $\phi_0$.  The integral (\ref{eqn:phicD'}) can be evaluated, with $\tau$-cutoffs at
$\delta$ and $L$ to give
\begin{align}
\label{eqn:phi0d2int}
\phi_0 &=  \frac{g z^{\dt}}2 \left[\int_{\delta/z}^{L/z} dy \left(1+(y-it_B/z)^2\right)^{-\frac12} 
+\text{c.c.} \right] \\
\label{eqn:gzd2G}
&= g z^{\dt} G(t_B/z, \delta/z, L/z),
\end{align}
where 
\beq \label{eqn:G}
G(s,\vep, l) = \frac12\left(\sinh^{-1} (l-is)-\sinh^{-1} (\vep-i s) + \text{c.c.} \right). 
\eeq
The dependence on $\delta$ in (\ref{eqn:gzd2G}) is needed only in the region $t_B\sim z$, 
everywhere else it may safely be taken to zero.  Also, since we will need this solution in the 
regions where $z$ and $t_B$ are at most on the order of $R\ll L$, we often use the limiting
form of this function taking $L\gg z,t_B$.  In particular, on the surface $\mathcal{E}$ with
$t_B=0$, it evaluates to
\beq\label{eqn:phi0Elog}
\phi_0\rightarrow g z^{\dt} \log{\frac{2L}{z}},
\eeq
plus terms suppressed by $\frac{z^2}{L^2}$. It is useful to express this in terms of the 
renormalized vev for $\op$ calculated in (\ref{eqn:vevOren}):
\beq
\phi_0\rightarrow -\vev{\op}^\text{ren.}_g z^{\dt} - g z^{\dt} \log\frac{\mu z}{2}.
\eeq
The $\log$ term in this expression is what would have resulted if we had cut the integral
(\ref{eqn:phi0d2int}) off at $\mu^{-1}$ rather than $L$.  
Finally, it is also useful to have the form of the function (\ref{eqn:gzd2G})
along $\mathcal{T}$, where $t_B\gg z$,
\beq\label{eqn:phi0Tlog}
\phi_0\rightarrow gz^{\dt} \log\frac{L}{t_B}.  
\eeq

At $t_B=0$, 
the solution $\phi_\omega$ is still given by a modified Bessel as in equation (\ref{eqn:phio}).
We also need expressions for the behavior of $\phi_\omega$ along the surface
$\mathcal{T}$.  When 
$t_B\ll \omega^{-1}$, the same arguments that led to equation (\ref{eqn:phiort}) for 
$\D<\dt$ can be applied to the defining integral for $\phi_\omega$ to show it takes
the form
\beq\label{eqn:phiobzlzG}
\phi_\omega = 
 \beta_\omega z^{\dt}+\lambda_\omega z^{\dt} G(t_B/z,\delta/z, a/z );
 \qquad \beta_\omega = -\gamma_E-\log\omega a,
\eeq
where $a$ is the intermediate scale introduced in the integral, as in equation (\ref{eqn:asplit}),
and satisfies $t_B
\ll a\ll \omega^{-1}$.  Note that this answer does not actually depend on $a$ since 
it will cancel between the $\log$ and $G$ terms, but it is convenient to make this 
separation when evaluating the $\mathcal{T}$ surface integrals in 
section \ref{sec:D=d2calc}.  From this, the form of $\phi_\omega$ can be read off for $t_B\gg z$:
\beq \label{eqn:phiologot}
\phi_\omega\rightarrow -\lambda_\omega z^{\dt} \left(\gamma_E + \log \omega t_B\right).
\eeq

\section{Surface integrals} 
This appendix gives the details of the $\mathcal{E}$ and $\mathcal{T}$ surface integrals
for $\D<\dt$ (section \ref{sec:D<dtcalc}) and for $\D=\dt$ (section \ref{sec:D=d2calc}).

\subsection{$\D<\dt$} \label{sec:D<dtcalc}
Each integral in this case will be proportional to one of $\vev{\op}_g\delta{\vev{\op}}$, 
$(g\delta\vev{\op}+\lambda(0)\vev{\op}_g)$, or $\lambda(0) g$.    In
each case, we show explicitly that the possibly divergent terms coming from the $z_0
\rightarrow 0$ limit cancel, as they must to give an unambiguous answer.

\paragraph{1. $\vev{\op}_g\, \delta\vev{\op} $ term.}
This term arises from the piece of $\phi_0$ and $\phi_\omega$ that goes like $\frac{-z^\D}{2\D
-d}$.  In particular, it has no dependence on $t_B$ anywhere.  On the surface $\mathcal{E}$, 
since $\partial_\tau\phi=0$, the integrand in (\ref{eqn:Eint}) 
only depends on $\nabla^2\phi^2$.  Working to leading
order in $R$ means only keeping the $z$ derivatives in the Laplacian.  The term in this 
expression with coefficient $\vev{\op}_g\,\delta\vev{\op}$ is $\frac{2 z^{2\D}}{(2\D-d)^2}$,
and acting with the Laplacian on this gives $\frac{4\D z^{2\D}}{2\D-d}$.  Then the $\mathcal{E}$
integral is
\begin{align}
\delta S^{(2)}_{\mathcal{E},1} &= -2 \pi \vev{\op}_g\,\delta\vev{\op}
\frac{\D \Omega_{d-2} }{2\D-d} \int_{z_0}^R dz\, z^{2\D-d-1}
\int_0^{\sqrt{R^2-z^2} }dr\, r^{d-2}\left[\frac{R^2-r^2-z^2}{2R}\right] \\
&=
\label{eqn:dSEz2D}
-2\pi \vev{\op}_g\,\delta\vev{\op}
\frac{\D\Omega_{d-2}}{d^2-1}\left[ R^{2\D}
\frac{\Gamma(\dt+\frac32)\Gamma(\D-\dt+1)}{ (2\D-d)^2 \Gamma(\D+\frac32)} 
- \frac{R^dz_0^{2\D-d} }{(2\D-d)^2}\right]. 
\end{align}
Note this consists of a finite term scaling as $R^{2\D}$ and a divergence in $z_0$.  

The divergence must cancel against the integral over $\mathcal{T}$, given by 
(\ref{eqn:Tint}).  Unlike the case $\D>\dt$, this integral has a vanishing contribution from
the region $t_B\sim z$, but instead a divergent contribution from $t_B\gg z$.  Again picking
out the $\vev{\op}_g\,\delta\vev{\op}$ term in the integrand (\ref{eqn:Tint}), we find
\begin{align}
\delta S^{(2)}_{\mathcal{T},1} &=  -2\pi 
\vev{\op}_g\,\delta\vev{\op} \frac{\Omega_{d-2} z_0^{-d+1}}{(2\D-d)^2}
\int_0^Rdt\int_0^{R-t}dr\,r^{d-2} \frac{t}{R} \left[2
(\D z_0^{\D-1})^2-\frac{\D}{z^2}(2\D-d)z^{2\D}\right]  \\
&=
-2\pi \vev{\op}_g\,\delta\vev{\op} \frac{\D \Omega_{d-2} R^d z_0^{2\D-d}  }{(d^2-1)(2\D-d)^2} .  
\end{align}
Here, we see this cancels the divergence in (\ref{eqn:dSEz2D}), and thus we are left with only 
the finite term in that expression.  

\paragraph{2. $g\delta\vev{\op}+ \lambda(0)\vev{\op}_g$ term. }
On the surface $\mathcal{E}$, this term comes from the part of one field going like $z^\D$, and 
the other going like $z^{d-\D}$.  Hence, when we evaluate this term in $\nabla^2\phi^2$ for the 
bulk integral, we will be acting on a term proportional to $z^d$, which is annihilated by the 
Laplacian.  So the bulk will only contribute terms that are subleading to $R^d$ terms from
$\delta S^{(1)}$. The calculation of these subleading terms would be similar to the 
calculation for in section \ref{sec:D>dt}, but we do not pursue this further here.

Instead, we examine the integral over $\mathcal{T}$, which can produce finite contributions.  
Along this surface, the fields are now time dependent, and hence all terms in equation
(\ref{eqn:Tint}) are important.  We start by focusing on the terms involving time derivatives
of $\phi$.  The $z$-derivative acts on the term going as $\frac{-z^\D}{2\D-d}$, 
and the $t$ derivative on
$z^{d-\D} F(t/z)$.  To properly account for the behavior of $F$ when $t\sim z$, it is useful
to split the $t$ integral into two regions, $(0,c)$ and $(c,R)$ with $z\ll c\ll R$.  In the 
first region this gives
\beq \label{eqn:0c}
-2\pi \frac{\D\Omega_{d-2}}{2\D-d} 
\int_0^c dt \int_0^R dr\, r^{d-2} \left(\frac{R^2-r^2}{2R}\right) \partial_t
F(t/z_0)  = \frac{-2\pi \D \Omega_{d-2} R^d}{(2\D-d)(d^2-1)} F(t/z_0)\Big|^{c}_0.
\eeq
From $(\ref{eqn:F})$, we see that $F(0)=1$, and the value at $t=c$ can be read off using the 
asymptotic form for $F$ in equation $(\ref{eqn:Fasym})$.  This form is also useful for 
evaluating the integral in the second region, where the integral is
\begin{align}
&\frac{-2\pi\D\Omega_{d-2}  (d-2\D)}{(2\D-d)} B
z_0^{2\D-d} \int_c^R dt\int_0^{R-t} dr\, r^{d-2}\left(\frac{R^2-r^2-t^2}{2R}\right)t^{d-2\D-1}
\nonumber \\
\label{eqn:z2D-d}
&= \frac{-2\pi\D\Omega_{d-2}  }{(2\D-d)}
Bz_0^{2\D-d} \left[ R^{2(d-\D)} \frac{d\,\Gamma(d-1)\Gamma(d-2\D+2)}{\Gamma(2d-2\D
+2)}   - \frac{c^{d-2\D} R^d }{d^2-1} \right],
\end{align}
where this equality holds for $c\ll R$. 
The second term cancels the $c$-dependent term of (\ref{eqn:0c}), while the first term is 
a remaining divergence which must cancel against the other piece of the $\mathcal{T}$ 
integral.  This is the piece coming from the second bracketed expression in equation
(\ref{eqn:Tint}).  This term receives no contribution from the region $t\sim z$, so we can
evaluate it  in the region $t\gg z$, using the asymptotic form for $F(t/z)$.  Evaluating 
the derivatives in this expression (and recalling that only the $z$-derivatives in the Laplacian
will produce a nonzero contribution at $z\rightarrow 0$), this leads to
\begin{align}
& \frac{2\pi\Omega_{d-2}}{(2\D-d)}B z_0^{2\D-d}\int_0^Rdt\int_0^{R-t} dr\, r^{d-2} 
\frac{d\D}{R} t^{d-2\D+1} \nonumber\\
&=
\frac{2\pi\D\Omega_{d-2}}{2\D-d} Bz_0^{2\D-d}\frac{d\,\Gamma(d-1) \Gamma(d-2\D+2)}{\Gamma(
2d-2\D+2)},
\end{align}
which cancels the remaining term in (\ref{eqn:z2D-d}).  

Hence the only contribution remaining comes from (\ref{eqn:0c}) at $t=0$, and gives
\beq\label{eqn:dS2T2}
\delta S^{(2)}_{\mathcal{T},2} = \frac{2\pi  \Omega_{d-2} R^d\D}{(d^2-1)(2\D-d)}(g\delta\vev{\op}
+\lambda(0)\vev{\op}_g).
\eeq

\paragraph{3. $g\lambda(0) $ term.} 
The final type of term arises when both fields behave as $z^{d-\D} F(t/z)$.  
The $\mathcal{E}$ surface term will go like $R^{2(d-\D)}$, and hence will be subleading 
compared to the $R^d$ terms.  In fact, this calculation is essentially the same as the 
change in vacuum entanglement when deforming by a constant source, and the form
of this term is given in equation (4.34) of \cite{Faulkner2015} (although 
that calculation was originally performed only for $\D>\dt$).  
Also there is no divergence in $z_0$ in these terms.

On the other hand, the integral over $\mathcal{T}$ does lead to potential divergences, but 
we will show that these all cancel out as expected.  We may focus on the region $t\gg z$ 
since there is no contribution from $t\sim z$.  Using the asymptotic form (\ref{eqn:Fasym})
for $F$, the part of the integral (\ref{eqn:Tint}) involving $t$ derivatives becomes
\begin{align}
&2\pi\Omega_{d-2} 2\D(d-2\D) B^2 z_0^{2\D-d} 
\int_0^Rdt\int_0^{R-t} dr\, r^{d-2} \left(\frac{R^2-r^2-t^2}{2R}
\right) t^{2d-4\D-1} \nonumber \\
&=
\label{eqn:R3d4D}
2\pi\Omega_{d-2} B^2 z_0^{2\D-d} R^{3d-4\D} 
\frac{\D d\,\Gamma(d-1) \Gamma(2d-4\D+2)}{\Gamma(3d-4\D+2)} .
\end{align}
Similarly, the second bracketed term in (\ref{eqn:Tint}) evaluates to 
\begin{align}
&-2\pi \Omega_{d-2} \D d B^2 z_0^{2\D-d} \int_0^Rdt\int_0^{R-t} dr\, r^{d-2} 
\frac{t^{2d-4\D+1}}{R} \nonumber \\
&= 
-2\pi \Omega_{d-2} B^2 z_0^{2\D-d} R^{3d-4\D} 
\frac{\D d\,\Gamma(d-1) \Gamma(2d-4\D+2)}{\Gamma(3d-4\D+2)} ,
\end{align}
perfectly canceling against (\ref{eqn:R3d4D}).  Hence, the $\mathcal{T}$ surface integral 
gives no contribution, and the full $g\lambda(0)$ contribution, coming entirely from the
$\mathcal{E}$ surface, is subleading.

\subsection{$\D=\dt$} \label{sec:D=d2calc}
Here we compute the surface integrals  and divergence in $\delta S^{(1)}$ 
when $\D=\dt$.  The calculation is divided into
four parts: the $\mathcal{E}$ surface integral, the $\mathcal{T}$ surface integral for 
$t_B\sim z_0$, the $\mathcal{T}$ surface integral for $t_B\gg z_0$, and the 
$\delta S^{(1)}$ divergence.

\paragraph{1. $\mathcal{E}$ surface integral.}
Equation (\ref{eqn:Eint}) shows that we need to compute the Laplacian acting on $(\phi_0
+\phi_\omega)^2$ .  At leading order, only the $z$-derivatives from the Laplacian contribute
since the other derivatives are suppressed by a factor of $z^2$.  Using the bulk solutions 
found for $\phi_0$ (\ref{eqn:phi0Elog}) and $\phi_\omega$ (\ref{eqn:phioElog}), the 
$\mathcal{E}$ surface integral at $O(\lambda^1 g^1)$ is
\begin{align}
\delta S^{(2)}_{\mathcal{E}} &= -4\pi \Omega_{d-2}  g\lambda_\omega
\int_{z_0}^R \frac{dz}{z} \int_0^{\sqrt{R^2-z^2}} dr\, r^{d-2} \left[\frac{R^2-r^2-z^2}{8R}\right]
\left[ 2 + d\gamma_E + d\log \frac{\omega z^2}{4L} \right] \nonumber \\
&=
-2\pi g\lambda_\omega\frac{\Omega_{d-2} R^d}{d^2-1}\int_{z_0/R}^1 \frac{dw}{w} 
(1-w^2)^{\frac{d+1}{2}}\left(1+\frac{d}{2}\gamma_E +\frac{d}{2} \log \frac{w^2 R^2\omega}{4L}
\right).
\end{align}
The divergence in $z_0$ comes from $w$ near zero, and so can be extracted by 
setting the $(1-w^2)$ term in the integrand to $1$, its value at $w=0$. The divergent integral
evaluates to
\beq \label{eqn:dS2Ediv}
\delta S^{(2)}_{\mathcal{E},\text{div.}} 
= -2\pi g\lambda_\omega\frac{\Omega_{d-2} R^d}{d^2-1}
\logp{\frac{R}{z_0}}\left(1+\frac{d}{2}\gamma_E+\frac{d}{2}\log\frac{\omega R z_0}{4L}\right),
\eeq
and the remaining finite piece with $z_0\rightarrow 0$ is 
\beq
\delta S^{(2)}_{\mathcal{E},\text{fin.}} 
= -2\pi g\lambda_\omega\frac{\Omega_{d-2} R^d}{d^2-1}
\int_0^1\frac{dw}{w} \left[(1-w^2)^{\frac{d+1}{2}} -1\right]\left(1+\frac{d}{2}\gamma_E+
\frac{d}{2}\log w^2\frac{R^2\omega}{4L}\right). \label{eqn:finitebulk}
\eeq
The following two identities are needed to evaluate this,
\begin{align}
&\int_0^1\frac{dw}{w}\left[(1-w^2)^{\frac{d+1}{2}}-1\right] = -\frac12 H_{\frac{d+1}{2}} 
\label{eqn:nologii}\\
&\int_0^1\frac{dw}{w}\left[(1-w^2)^{\frac{d+1}{2}}-1\right]\log w = 
\frac18\left(H_{\frac{d+1}{2}}^{(2)} + H_{\frac{d+1}{2}}^2\right), \label{eqn:logii}
\end{align}
where the harmonic number $H_n$ was defined below equation (\ref{eqn:Deqdt}), and 
$H^{(2)}_n$ is a second order harmonic number, defined for the integers by $H^{(2)}_n = 
\sum_{k=1}^n \frac{1}{k^2}$, and  for arbitrary complex $n$ by $H^{(2)}_n = \frac{\pi^2}{6}
-\psi_1(n+1)$, where $\psi_1=\frac{d^2}{dx^2}\log \Gamma(x)$.  
With these, the finite piece (\ref{eqn:finitebulk}) becomes
\beq\label{eqn:dS2Efinlog}
\delta S^{(2)}_{\mathcal{E},\text{fin.}}  = 2\pi g\lambda_\omega\frac{\Omega_{d-2} R^d}{d^2-1}
\left[\frac{d}{4}H_{\frac{d+1}{2}}\left(\gamma_E + \log\frac{\omega R^2}{4L}\right)
-\frac18\left(H_{\frac{d+1}{2}}^{(2)} + H_{\frac{d+1}{2}}(H_{\frac{d+1}{2}}-2)\right) \right].
\eeq

\paragraph{2. $\mathcal{T}$ surface near $t_B\sim z$.}
This region contains several divergences in $z_0$ and $\delta$.  
The specific range of $t_B$ will be $t_B\in(0,c)$, with $z\ll c\ll R$.  Only the first 
bracketed term in (\ref{eqn:Tint}) contributes in this region, and using the general 
solutions for $\phi_0$ and $\phi_\omega$ from equations (\ref{eqn:gzd2G}) and 
(\ref{eqn:phiobzlzG}), it gives at $O(\lambda^1 g^1)$
\beq \label{eqn:dS2TG}
\delta S^{(2)}_{\mathcal{T},\text{div.}} = 2\pi g \frac{\Omega_{d-2} R^d}{d^2-1} \int_0^c dt \left[
\dt\partial_t\left(\lambda_\omega G_L G_a + \beta_\omega G_L\right)
+\lambda_\omega z_0\left(\partial_z G_L\partial_t G_a + \partial_z G_a\partial_t G_L\right)
\right],
\eeq
having introduced the shorthand $G_L\equiv G(t/z_0,\delta/z_0,L/z_0)$ 
and similarly for $G_a$.  The
first term in this expression is a total derivative so can be integrated directly.  The boundary
term at $t=0$ is
\beq\label{eqn:Tz0div}
 2\pi g \lambda_\omega 
\frac{\Omega_{d-2} R^d}{d^2-1} \dt \logp{\frac{2L}{z_0}} \left(\gamma_E+
\log{\frac{\omega z_0}{2}}
\right).
\eeq
At the other boundary $t=c\gg z_0$, the asymptotic formulas (\ref{eqn:phiologot}) and 
(\ref{eqn:phi0Tlog})  produce the term 
\beq\label{eqn:Tcdiv}
-2\pi g \lambda_\omega 
\frac{\Omega_{d-2} R^d}{d^2-1} \dt\logp{\frac{L}{c}} \left(\gamma_E+\log\omega c\right).
\eeq

The remaining terms in (\ref{eqn:dS2TG}) contain a divergence in $\delta$, coming from $t
\sim z$.  To extract it, we focus specifically on the regions $(z_0-u, z_0+v)$ and $(z_0+v, c)$,
 where 
$u,v\ll z$ and positive.  It is straightforward to show that the integral over the region $(0,z_0-u)$ 
is $O(\delta)$, and so does not contribute when $\delta$ is sent to zero.  The divergence 
in the $(z_0-u,z_0+v)$ region can be evaluated by taking a scaling limit with a 
change of  variables, $t_B
= z_0 + s \delta$, and expanding the integrand about $\delta =0$.  After also taking the limit 
$L/z_0, a/z_0\rightarrow \infty$ in the integrand, the integral in this region becomes
\beq \label{eqn:Tddiv}
-\lambda_\omega \int_{-u/\delta}^{v/\delta} ds\, \frac{s+\sqrt{1+s^2}}{1+s^2} 
\rightarrow-\lambda_\omega \log\frac{2v}{\delta},
\eeq
which holds for $u,v\gg\delta$.  For the region $(z+v,c)$, we can take $\delta/z\rightarrow 0$
and $L/z, a/z\rightarrow\infty$, which produces the integral
\beq\label{eqn:Tddivct}
2 \lambda_\omega \int_{z_0+v}^c dt \left(\frac{1}{\sqrt{t^2-z_0^2}} - \frac{t}{t^2-z_0^2}\right)
\rightarrow \lambda_\omega \log\frac{8v}{z_0},
\eeq
where we have taken the limits $c/z_0\gg1$, $v/z_0\ll 1$.  

The final collection of the four contributions (\ref{eqn:Tz0div}), (\ref{eqn:Tcdiv}), 
(\ref{eqn:Tddiv}) and (\ref{eqn:Tddivct})  is 
\begin{align}
\delta S^{(2)}_{\mathcal{T},\text{div.}} = 2\pi g \lambda_\omega \frac{\Omega_{d-2}R^d}{d^2-1}
\left[\dt \logp{\frac{2L}{z_0}}\left(\gamma_E+\log\frac{\omega z_0}{2}\right)
-\dt\logp{\frac{L}{c}}\left(\gamma_E+\log\omega c\right) 
+ \log \frac{4\delta}{z_0}
\right].
\label{eqn:dS2Tdivlog}
\end{align}

\paragraph{3. $\mathcal{T} $ surface for $t_B\gg z$. }
In this region, $t_B\gg z$, and we can use the asymptotic forms (\ref{eqn:phi0Tlog}) 
and (\ref{eqn:phiologot})
for the fields $\phi_0$ and $\phi_\omega$. We start with the first bracketed term in
equation (\ref{eqn:Tint}),
\begin{align}
\delta S^{(2)}_{\mathcal{T},1} &= 
2\pi g\lambda_\omega \Omega_{d-2} \int_c^R dt\int_0^{R-t} dr\, r^{d-2} 
\left[\frac{R^2-r^2-t^2}{2R}\right] \frac{d}{2t} \left(\gamma_E+ \log\frac{t^2\omega}{L}\right) \\
&= 2\pi g \lambda_\omega\frac{\Omega_{d-2} R^d}{d^2-1} \frac{d}{2} 
\int_{c/R}^1 \frac{ds}{s}(1-s)^d(1+ds) \left(\gamma_E+\log\frac{s^2 R^2\omega}{L}\right).
\end{align}
The divergence in this integral comes from $s=0$, so it can be separated out by setting 
$(1-s)^d(1+ds)$ to $1$ (its value at $s=0$), leading to 
\beq\label{eqn:logRcdiv}
\int_{c/R}^1\frac{ds}{s}\left( \gamma_E + \log\frac{s^2 R^2\omega}{L}\right) = 
\log\left(\frac{R}{c}\right)\left(\gamma_E+ \log\frac{cR\omega}{L}\right).
\eeq
The remaining finite piece of the integral is
\beq\label{eqn:remfin} 
\int_0^1\frac{ds}{s} \left[(1-s)^d(1+ds)-1\right]\left(\gamma_E+\log\frac{s^2 R^2 \omega}{L}\right).
\eeq
Evaluation of this integral involves the following identites,
\begin{align}
&\int_0^1 \frac{ds}{s} \left[(1-s)^d(1+ds)-1\right] = 1-H_{d+1},\\
&\int_0^1\frac{ds}{s}\left[(1-s)^d(1+ds)-1\right]\log s=\frac12\left(H^{(2)}_{d+1} +H_{d+1}
(H_{d+1}-2)\right),
\end{align}
where the harmonic numbers $H_n$ and $H_n^{(2)}$ were defined below equations 
(\ref{eqn:Deqdt}) and (\ref{eqn:logii}). Using these to compute (\ref{eqn:remfin}), and 
combining the answer with equation (\ref{eqn:logRcdiv}) gives
\begin{align}
\delta S^{(2)}_{\mathcal{T},1} =
2\pi g \lambda_\omega& \frac{\Omega_{d-2} R^d}{d^2-1}\dt\left[ 
\logp{\frac{R}{c}}\left(\gamma_E+\log\frac{cR\omega}{L}\right)   \right. \nonumber \\
& \label{eqn:dS2T2log}
 \left. -(H_{d+1}-1)\left(\gamma_E+\log\frac{R^2\omega}{L}\right)
+H^{(2)}_{d+1} +H_{d+1}(H_{d+1}-2)\right].
\end{align}

Finally, we compute the second bracketed term of (\ref{eqn:Tint}).  Only the $z$-derivatives 
in the Laplacian term $\nabla^2 \phi^2$  
contribute in the limit $z\rightarrow 0$. Since $\phi^2$ scales as $z^d$, 
the $z$-derivatives in the Laplacian annihilate it, and hence this piece is zero.  
The integral then becomes
\begin{align}
\delta S^{(2)}_{\mathcal{T},2}&=2\pi g\lambda_\omega \Omega_{d-2}\left(\dt\right)^2 
2 \int_0^R dt \int_0^{R-t} dr r^{d-2}
\frac{t}{R}\logp{\frac{L}{t} } (\gamma_E+\log\omega{t}) \\
&= 2\pi g \lambda_\omega  \frac{\Omega_{d-2} R^d}{d^2-1}  \frac{d}{2} 
\left[ 
-H_{d+1}^{(2)} -H_{d+1}(H_{d+1}-2) +(H_{d+1}-1)\left(\gamma_E +\log\frac{R^2\omega}{L}
\right) 
\right. \nonumber \\
&\hphantom{=2\pi g \lambda_\omega  \frac{\Omega_{d-2} R^d}{d^2-1}  \frac{d}{2} } \left.
-\logp{\frac{R}{L}} \left(\gamma_E+\log R\omega\right)
\right].
\end{align}
The finite terms cancel against those appearing in (\ref{eqn:dS2T2log}), and the final
combined result is 
\beq \label{eqn:dS2T1+2}
\delta S^{(2)}_{\mathcal{T},1+2} = 
2\pi g \lambda_\omega \frac{\Omega_{d-2} R^d}{d^2-1}\dt
\logp{\frac{L}{c}} \left(\gamma_E+\log\omega c\right),
\eeq
which perfectly cancels the $c$-dependent terms in (\ref{eqn:dS2Tdivlog}).  Hence, 
no finite terms result from the integral along $\mathcal{T}$ in the $t_B\gg z$ region.

\paragraph{4. $\delta S^{(1)}$ term. }
The final divergence in $\delta$ comes from the expectation value of the CFT stress tensor,
in $\delta S^{(1)}$.  At order $g \lambda_\omega$, this is given by
\beq
\delta\fvev{T^0_{00}(0)}=  -\int d^d x_a d^d x_b g\lambda_\omega(x_b)
\fvev{T^0_{\tau\tau}(0) \op(x_a)\op(x_b)}.
\eeq
The only divergence in this correlation function comes from when $x_a
\rightarrow x_b\rightarrow 0$, and is logarithmic in the cutoff  $\delta$.  As was the
case for the logarithmic divergence in $\vev{\op}$, regulating this divergence involves 
introducing a renormalization scale $\mu$ that separates the divergence from the 
finite part of the correlation function.  This is done by cutting off the $\tau$ integrals when 
$|\tau_a|\geq\mu^{-1}$ and 
$|\tau_b|\geq\mu^{-1}$.  

The divergence comes from the leading piece in the expansion of $\lambda_\omega(x)$
about $x=0$, 
\beq
\delta\fvev{T^0_{\tau\tau}(0) }_{\text{div.}} =  g \lambda_\omega \int d^d x_a d^d x_b 
\fvev{T^0_{\tau\tau}(0) \op(x_a)\op(x_b)}.
\eeq
This divergence can be evaluated using the same method described in Appendix D of 
\cite{Faulkner2015}.  The translation invariance of the correlation function allows one to 
write it as an integral of the stress tensor averaged over the spatial volume,
\beq
g\lambda_\omega \frac1V \int d^{d-1}\vec{x}
\int_{C(\delta,\mu)} d\tau_a \int_{C(\delta,\mu)} d\tau_b \int d\vec{x}_a d\vec{x}_b
\fvev{T^0_{\tau\tau}(0,\vec{x}) \op(x_a)\op(x_b)}.
\eeq

The stress tensor integrated over $\vec{x}$ is now a conserved quantity, and so the surface 
of integration may deformed away from $\tau=0$.  As long as it does note encounter 
the points $\tau_a$
or $\tau_b$, the surface can be pushed to infinity, so that the correlation function vanishes.  
This is possible if $\tau_a$ and $\tau_b$ have the same sign.  However, when $\tau_a$ and 
$\tau_b$ have opposite signs, one of them will be passed as  the surface is pushed to
infinity.  This leads to a contribution from the operator insertion at that point, as dictated by 
the translation Ward identity.  Let us choose to push past $\tau_a$.  For $\tau_a<0$,
the contribution from the operator insertion is 
\begin{align} 
&\hphantom{=}
-g\lambda_\omega\frac1V \int d\vec{x} d\vec{x_a} d\vec{x_b}\int_\delta^\mu d\tau_b
\int_{-\mu}^{-\delta} d\tau_a \partial_{\tau_a}\fvev{\op(x_a)\op(x_b)} \delta(\vec{x}-\vec{x_a}) \\
&=  -g\lambda_\omega c_\Delta' S_{d-2} \frac{\sqrt{\pi}\,\Gamma(\dt-\frac12)}{2\Gamma(\dt)}
\int_\delta^\mu d\tau_b \left[\frac1{\tau_b+\delta} - \frac1{\tau_b +\mu}\right] \\
&=-\frac12g\lambda_\omega \log{\frac{\mu}{4\delta}},
\end{align}
where in this last equality we have taken $\mu\gg\delta$. 
It is straightforward to check that for $x_a^0>0$, you get the same contribution, 
so that the full divergent
piece of the stress tensor is 
\beq
\delta\fvev{T_{00}(\vec{x})}_{\text{div.}} = 
g\lambda_\omega \log{\frac\mu{4\delta}}.
\eeq
This then defines a renormalized stress tensor expectation value,
\beq
\delta\vev{T_{00}(0)}^{\text{ren.}} = \delta\vev{T_{00}(0)} - g\lambda_\omega\log\frac{\mu}{4\delta}
\eeq

Finally, the contribution to $\delta S^{(1)}$ comes from integrating $\delta\vev{T_{00}(\vec{x})}$
over the ball $\Sigma$ according to equation (\ref{eqn:dS1}).  Since the stress tensor 
expectation value may be assumed constant over a small enough ball, the expression
for $\delta S^{(1)}$ in terms of the renormalized stress tensor expectation value is 
\beq \label{eqn:deltact}
\delta S^{(1)}_{\lambda g} =
2\pi \frac{\Omega_{d-2} R^d}{d^2-1}
\left(\delta\vev{T^0_{00}}^{\text{ren.}} + g\lambda_\omega \logp{\frac{\mu}{4\delta}}  \right).
\eeq

\bibliographystyle{JHEP}
\bibliography{diamondEE}

\end{document}